\documentclass[vecphys]{svmult}

\usepackage{makeidx}         % allows index generation
\usepackage{graphicx}        % standard LaTeX graphics tool
\usepackage{multicol}        % used for the two-column index
\usepackage[bottom]{footmisc}% places footnotes at page bottom
\makeindex             % used for the subject index
\def\result{${\cal R}_{\rm BGV}$}
\def\edge{${\cal E}$}
\def\boundary{${\cal B}$}
\def\hotpocket{hotzone}
\def\gsim{\;\rlap{\lower 2.5pt
 \hbox{$\sim$}}\raise 1.5pt\hbox{$>$}\;}
\def\lsim{\;\rlap{\lower 2.5pt
   \hbox{$\sim$}}\raise 1.5pt\hbox{$<$}\;}

\begin{document}

\title*{Eternal Inflation, past and future}
\author{Anthony Aguirre\inst{1}}
\institute{Department of Physics/SCIPP\\ UC Santa Cruz\\ 1156 High St., Santa Cruz, CA 95064 \\
\texttt{aguirre@scipp.ucsc.edu}}
\maketitle

Cosmological inflation, if it occurred, radically alters the picture of the `big bang', which would merely point to reheating at the end of inflation.  Moreover, this reheating may be only {\em local} so that inflation continues elsewhere and forever, continually spawning big-bang-like regions.  This chapter reviews this idea of `eternal inflation', then focuses on what this may mean for the ultimate beginning of the universe.  In particular, I will argue that given eternal inflation, the universe may be free of a cosmological initial singularity, might be eternal (and eternally inflating) to the past, and might obey an interesting sort of cosmological time-symmetry.

\section{The inflationary Genie}

Cosmological inflation, as discussed at length in Chapter 2, was developed as a means of explaining the very simple yet specific `initial' conditions that define the hot big-bang cosmological model.  But while inflation grants the wish of providing a very large, uniform, hot, monopole-free region with appropriate density fluctuations, it is also, like the proverbial Genie let out of the bottle, difficult to contain.  In most inflation models, the selfsame accelerated expansion that allows inflation's explanatory and predictive successes also completely changes the ultra-large-scale structure of the universe in which it occurs, through the process of `everlasting' or `eternal' inflation.\footnote{Using `everlasting' inflation to refer to inflation that continues indefinitely into the future only, while reserving `eternal' inflation for inflation that also has always {\em been} occurring would probably be the best nomenclature.  But because the conventions are well-entrenched, I shall use `everlasting' and `eternal' inflation interchangeably for the former scenario, and `past-eternal inflation' for the latter.}

In this picture, once inflation starts, it continues forever -- at least somewhere -- continually spawning non-inflating regions large enough to be our entire observable universe.  While studiously ignored by many cosmologists (who would like to view inflation as a brief interlude between an initial singularity and the classic big-bang cosmology), eternal inflation has been a central preoccupation of many of inflation's inventors from the beginning, and appreciation of its profound implications for cosmology have steadily been spreading as inflation passes more and more observational tests.

Probably the deepest of these implications is that in inflation, the creation of an ultrahot, ultradense, nearly-uniform expanding medium -- previously equated with the `big-bang' and considered inextricably tied to the beginning of the universe -- now simply represents the {\em end} of inflation.  This moves the issue of the universe's beginning (if there was one!) into a wider context of a rather different character.  That shift, and what it may mean, will be the primary subject of this chapter.  I first review, in Sec.~\ref{sec-eireview}, what exactly eternal inflation is, developing in detail a particularly simple and rigorously understood realization of the idea, then discussing it in more general inflationary scenarios; in this way it will become clear how the inflationary background both houses and connects with hot, homogeneous expanding regions that could desribe our observable universe. (Hereafter, I will refer to such regions as `\hotpocket s', despite a strong temptation -- for reasons evident below -- to call them HotPocket$^{\textregistered}$s.) In Sec.~\ref{sec-obsprobs} I then discuss the significant and perhaps disturbing implications of this new picture for {\em testing} cosmological models.  

Most of the classic singularity theorems of Steven Hawking, Roger Penrose and others -- which underlie the widespread conviction in an initial cosmological  singularity -- simply do not apply to an inflating spacetime, or apply with considerable subtlety (see Ch.~[ellis] for more on this).  Thus inflation and eternal inflation enable, even require, a re-examination of this issue, and this will occupy the remainder of the chapter.  First, in Sec.~\ref{sec-startinginf} I review what is known about the possibility of {\em starting} inflation from a non-inflationary region.  This both sheds some light on the issue of inflation's beginning, and also lays some groundwork for better understanding general eternally-inflating models.  Next, in Sec.~\ref{sec-singth} I review singularity theorems that do say something about inflating spacetimes.  I focus in some depth on the results of~\cite{Borde:2001nh}, which are widely cited as showing that inflation cannot be eternal to the past. I think this inference is incorrect, and will explain why by discussing what I think it means to be past-eternal, and explicating models that fit this definition in Sec.~\ref{sec-pastei}.  Nonetheless, these singularity theorems may point to something very interesting about the arrow of time; in Sec.~\ref{sec-nonsing} I expand these thoughts into a discussion about the more general -- but related -- question of whether an inflationary cosmology must contain a `cosmological singularity', i.e. a beginning of classical time. 

\section{Everlasting inflation}
\label{sec-eireview}

The essence of eternal inflation is that while inflation eventually ends at any given location, the exponential creation of new volume allows inflation to continue forever globally:  inflation ends everywhere, yet goes on forever!  This peculiar behavior can be seen perhaps most clearly, and with the most mathematical control, in the situation of eternal inflation as driven by a `false-vacuum'.

\subsection{False-vacuum-driven eternal inflation}
\label{sec-fvei}

\begin{figure}
\centering
\includegraphics[width=8cm]{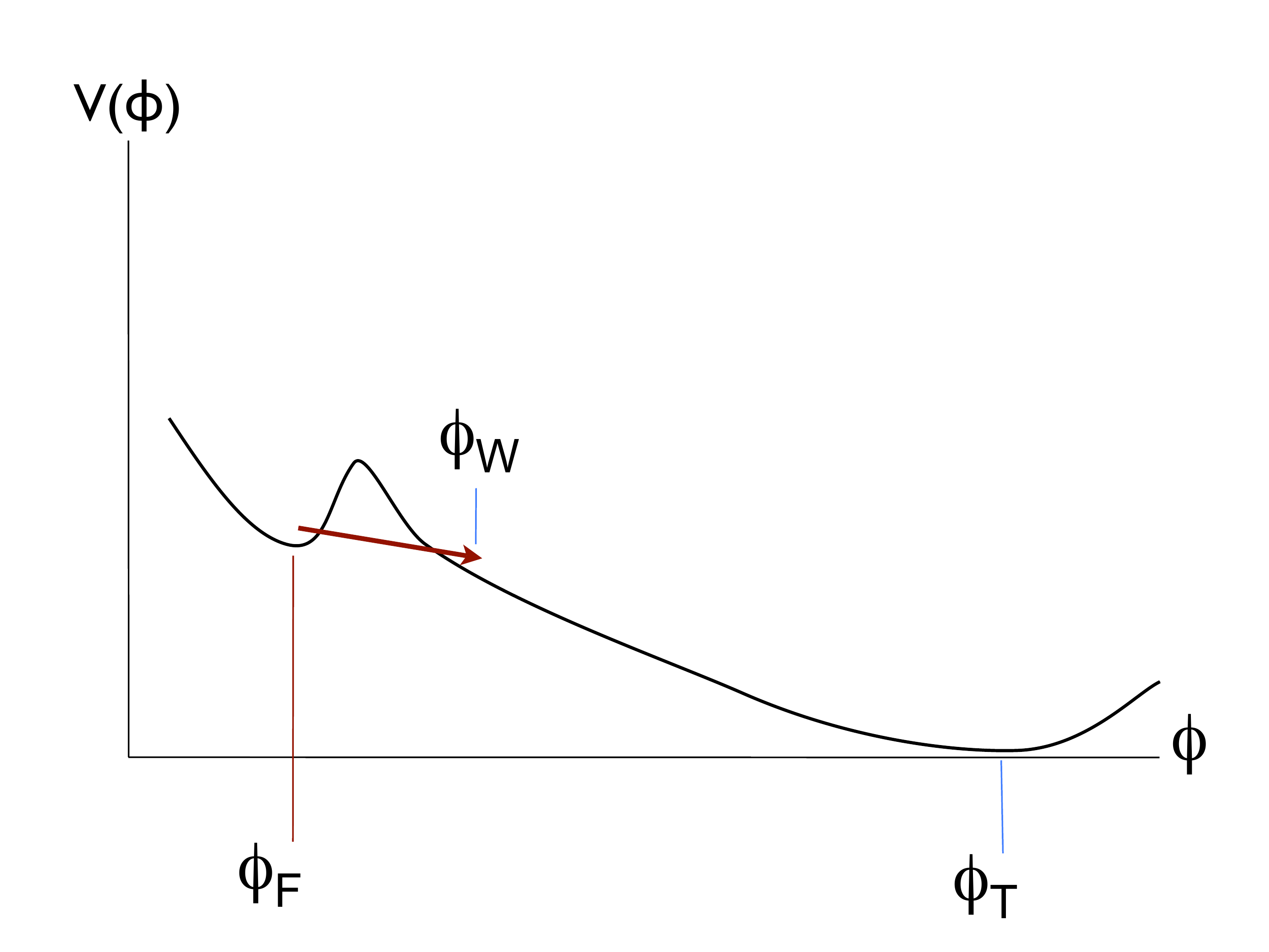}
\caption{A double-well inflationary potential $V(\phi)$.  Here, $V(\phi_F)$ corresponds to an inflationary vacuum energy, and $V(\phi_T)$ corresponds to our observed vacuum energy. While classically the field could never escape from near $\phi_F$, quantum tunnelling allows the formation of a region of $\phi_W$.  Subsequent inflation and reheating may occur as the field then rolls to the true minimum at $\phi_T$.
\label{fig-doublewell}}
\end{figure}

As a simple toy-model, consider the inflaton potential depicted in Fig.~\ref{fig-doublewell}. The global minimum is tiny and positive, at a field value $\phi_T$, but there is also a local or `false' minimum or `vacuum' at $\phi=\phi_F$, and we imagine that $V(\phi_F)$ corresponds to an inflationary vacuum energy -- say $\sim (10^{14}\,{\rm GeV})^4$ (or $\sim 10^{73}\,$g\,cm$^{-3}$ just to be clear what sort of numbers we are talking about).  

Now suppose that in some manner, a region of spacetime is produced which is locally approximately describable using the FRW metric with time-variable $t$, and has $\phi \approx \phi_F$ and $(1/\phi)(\partial\phi/\partial t) \ll H_F$ in a region of scale $\gg H_F^{-1}$, where $H_F = \sqrt{8\pi G V(\phi_F)/3}$ is the inflationary Hubble parameter, and $G$ is Newton's constant.  Then, as discussed in Ch.~2, this region will inflate with the scale factor $a(t) \propto \exp(H_Ft)$, so that the physical volume in some spatial region of constant size in comoving coordinates has a physical volume that increases as $\exp(3H_Ft)$.

What happens then?  In purely classical field theory, nothing.  The equations of motion governing the scalar field are much like those of a ball in the corresponding gravitational potential. Thus the ball would sit at rest in the false minimum, and the universe inflate,   forever. The region surrounding any point in the original patch would very quickly approach de Sitter spacetime (hereafter `dS'), as described by the metric
\begin{equation}
ds^2=-c^2 dt^2+\exp(2Ht)\left[dr^2+r^2d\Omega^2\right]
\label{eq-flatfoliation}
\end{equation}
(see Fig.~\ref{fig-dsreview} for some subtleties). 

Quantum mechanics makes things more interesting in two ways.  First, as first detailed by Gibbons \& Hawking~\cite{Gibbons:1977mu}, quantum fields in de Sitter spacetime are accurately described as {\em thermal}, with a temperature $T_{\rm GH}=\hbar H_F/2\pi k_B$ (where $\hbar$ and $k_B$ are Planck's and Boltzmann's constants).  Thus thermal fluctuations in the energy of the field can in principle allow it to hop over the barrier.  Second, just as for a classical particle-in-well, the field can tunnel through the barrier.  Through a combination of both processes, a small bubble of a new phase of field value $\phi_W$ can form, and expand as it converts volume from high- to low-vacuum energy and feeds the liberated energy into the kinetic energy of the bubble wall.

\begin{figure}
\centering
\includegraphics[width=12cm]{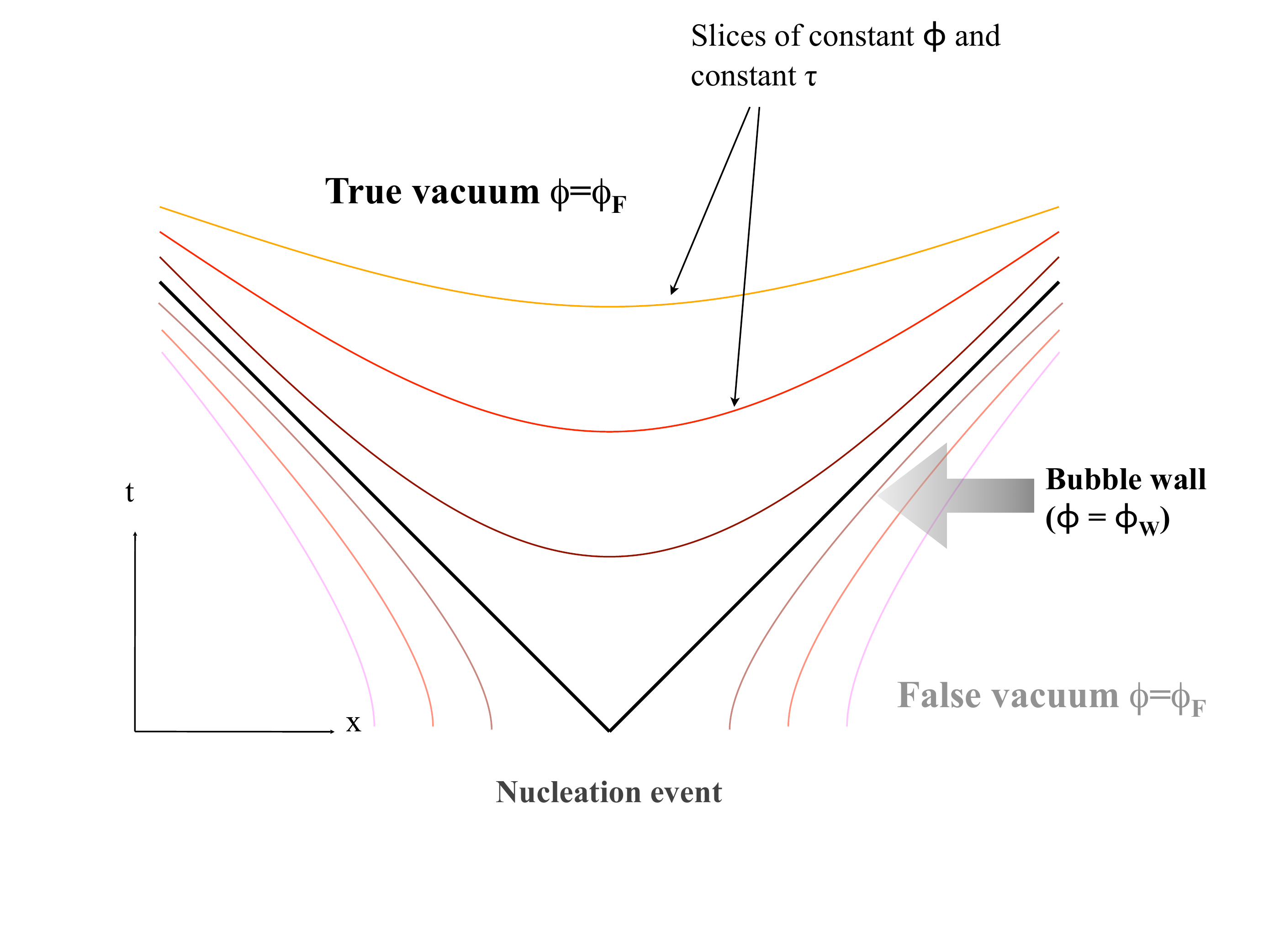}
\caption{The geometry of a nucleated bubble that could describe our observable universe back to very early times.  Lines correspond to surfaces of constant field value $\phi$. From the nucleation event, a surface of $\phi=\phi_W$ (the `tunneled-to' field value) expands at the speed of light.  Nestled into this light-cone are constant-$\phi$ hyperboloids, each of which has the geometry of an infinite negatively-curved homogenous space.  The sequence of nested hyperboloids corresponds to the time-sequence during which the inflaton rolls down the hill toward $\phi_T$ in Fig.~\ref{fig-doublewell}, and then to constant times in a big-bang universe inside, including reheating, recombination, and the present time.
\label{fig-bubble}}
\end{figure}

This process was described beautifully and fairly comprehensively in flat-space scalar field theory by Coleman~\cite{1977PhRvD..15.2929C}, then including gravity by Coleman \& De Luccia (hereafter `CDL')~\cite{Coleman:1980aw}.  There, they showed that the path integral describing the amplitude for tunneling could, by an analytic continuation, be equated with the integral over an `instanton', which is a regular solution to the Euclidean Einstein and scalar field equations satisfying appropriate boundary conditions.  
Moreover, the analytic continuation of this instanton back to a Lorentzian spacetime describes the structure of a bubble of true vacuum, and is depicted in Fig.~\ref{fig-bubble}.  

The particular field value $\phi_W$ to which the field tunnels holds on the forward light-cone of some point that we can denote the `nucleation point'.  Outside of this lightcone are (timelike) surfaces of constant field describing the bubble wall, with field values between $\phi_W$ and $\phi_F$ (far from the lightcone).  In certain forms of the potential, and long after the nucleation event, the wall can be considered very thin, and to coincide with the $\phi=\phi_W$ surface so that we can think of the wall as an expanding null-cone~\cite{Bucher:1994gb,Coleman:1980aw}. 

Inside the light-cone, spacelike surfaces of constant $\phi$ turn out, very interestingly, to also constitute (infinite) surfaces of constant negative curvature.  Therefore, if these homogeneous surfaces are taken to be constant-time surfaces (as general relativity leaves us free to do), then they describe an open Friedmann-Robertson-Walker cosmology nestled inside the forward light cone, with metric
\begin{equation}
ds^2=-c^2d\tau^2+a(\tau)^2 \left[d\xi^2+\sinh^2\xi d\Omega^2\right].
\label{eq-openflrw}
\end{equation}
The structure of a CDL bubble then looks entirely different depending upon how spacetime is `foliated' into space and time.  In the foliation outside that gives metric~\ref{eq-flatfoliation}, the bubble is finite, non-uniform, and growing; in the foliation that gives metric~\ref{eq-openflrw}, it is infinite and homogeneous.

While any foliation is equally valid in principle, inside the bubbles the homogeneous foliation is far more appropriate than others.  In particular, as the field rolls from $\phi_W$ toward $\phi_T$, inflation can continue (if the slow-roll conditions of Ch. 2 are met), and if this rolling continues for many e-foldings, the scale of the negative curvature in the bubble can become macroscopically large.  Then, once slow-roll fails and reheating occurs (assuming $\phi$ is coupled to other fields), the universe looks essentially like an infinite, nearly-flat, uniform \hotpocket\ that could describe our observed universe.  (There is in fact a large literature on such `open inflation' models as candidates for our observable universe; see, e.g. \cite{Gott:1982zf,Bucher:1994gb,Linde:1995rv,Yamamoto:1995sw,Linde:1998iw,Garriga:1998he,Linde:1999wv}.) Figure~\ref{fig-bubble} is worth careful study since (as discussed briefly below) it may well represent the current best bet for how the observable universe actually originated.

Now let us return to the background inflating space.  In a usual first-order phase transition, bubbles nucleate, expand, overlap, and percolate so that the phase transition completes and releases a certain amount of latent heat (depending upon the details of the potential).  Here, the situation is different due to the exponential expansion of the spacetime. Suppose we take a region of fixed comoving volume $V$ that has physical volume ${\cal V}=V \gg H^{-3}$ at some time $t=t_0$ in metric~\ref{eq-flatfoliation} (i.e. $a(t_0)=1$).  How much remains inflating as time goes on?  Because bubbles are (approximately) null, a point is still inflating if no bubble nucleated in its past lightcone, and the expected number of bubbles that would have passed through that point is given by the the 4-volume $v_4(t,t_0)$ of that past lightcone times the bubble nucleation rate $\lambda$ per physical time per physical volume.  It can then be shown (e.g.,\cite{Guth:1981uk}) that the inflating {\em fraction} is $f_{\rm inf} \equiv  {\cal V}_{\rm inf}/{\cal V} = V_{\rm inf}/V = \exp [-\lambda v_4(t,t_0)]$.  It can further be shown that in coordinates giving metric~\ref{eq-flatfoliation}, we have $v_4= 4\pi(t-t_0)/3H^3$.  This implies that $f_{\rm inf} \rightarrow 0$ as $t\rightarrow \infty$.  However, this just reduces the {\em comoving} inflating volume (where $V$ is fixed).  The {\em physical} inflating volume is $${\cal V}_{\rm inf}= \exp[3H(t-t_0)]f_{\rm inf}V \propto \exp[(3H-4\pi\lambda/3H^3)(t-t_0)],$$
which {\em increases} exponentially for $\lambda < 9H^{4}/4\pi$.

Thus as promised, for small $\lambda$, inflation never ends globally, even though a point at any given coordinates {\em will} (with probability one) eventually experience the end of inflation.

\subsection{Other modes of eternal inflation}

While the {\em way} in which inflation is eternal is clear when a local minimum occurs in the inflaton potential, eternal inflation also occurs in connection with other potential features.

First, consider local maxima of the inflaton potential. A region with the field originally at this maximum $\phi=\phi_M$ is unstable toward evolution toward larger or smaller field values.  But for a sufficiently large region this cannot happen coherently and thus a topological defect (a domain wall for a single scalar field model) will remain at the boundary between domains. If the potential is flat enough near $\phi_M$, this region can be both dominated by vacuum energy and also larger than $H_M^{-1} = \sqrt{3/8\pi G V(\phi_M)}$, and thus inflating (e.g.,\cite{Linde:2005ht}).  Due to the topological obstruction, the inflating region can never be removed; but away from it the field can eventually reach regions of the potential where slow-roll fails and reheating occurs, creating `pocket universes'. (I will reserve the term `bubble universe' for regions formed via bubble nucleation; `pocket universe' applies in general to a \hotpocket\ embedded in an eternally inflating background). This `topological eternal inflation' occurs in, e.g., `new inflation'~\cite{Linde:1981mu,Vilenkin:1983xq}, one of the earliest inflation models, but like false-vacuum eternal inflation is much more general. See, e.g.~\cite{Linde:1994wt} for a more detailed treatment.

Second, eternal inflation can occur when the potential is sufficiently flat and the field value sufficiently large.  In particular, it is generally argued that if, during an inflationary hubble time $H^{-1}$, the quantum fluctuations in the field -- which are of order $\delta\phi \sim T_{\rm GH} \propto \sqrt{V(\phi)}$ -- are larger than the classical rolling of the field, then eternal inflation ensues.  While it is not presently possible to do as rigorous a demonstration of this as in the false-vacuum or topological cases, it is clearly analogous:  consider an interval of field value $[\phi_E,\phi_E+\delta\phi]$ where the `eternality' condition holds, and a comoving region of size $> H^{-1}$ over which these values hold. Then, in an e-folding time $H^{-1}$, 
  the region's physical volume increases by $e^3$, so as long as more than $1/e$ of the comoving volume maintains field values in the interval, regions with these field values will exponentially increase in physical volume forever, just as in the false-vacuum case.
This version of eternal inflation is slightly younger than the others~\cite{Linde:1986fc}, but nevertheless like them almost as old as the idea of inflation itself.

\subsection{How generic is eternal inflation?}

Because eternal inflation can occur where the inflaton potential is at a minimum, maximum or neither, it is worth asking how generic eternal inflation is.  In particular, we might ask two questions:
\begin{enumerate}
\item Given some measure over inflaton potentials and starting field configurations, what is the measure over the combined space that gives both correct predictions of the currently-known inflationary observables and {\em also} is eternal, versus non-eternal?  This would, in some sense, tell us whether it is reasonable (i.e. not fine-tuned) to think that inflation might account for the observed universe without implying the existence of other `universes'.
\item Given some potential minimum (where the inflaton field presumably lies locally), we can expand the inflaton field about this minimum and express the observables in terms of the expansion coefficients.  We could also express the conditions of eternality in these -- what are the links?
\end{enumerate}

Unfortunately, neither question has really been investigated in the literature, and we have only fairly hand-wavy and qualitative arguments for eternal inflation being generic.

The first is historical: of the inflation models that have been proposed, most single-field models appear to be eternal according to the criteria outlined above.   (Moving to more fields, however, might well alter this conclusion, as for example two-field ``hybrid" inflation models (e.g.~\cite{Linde:1993cn}) are {\em not} eternal.)
The second argument is more general and intuitive: insofar as the {\em ending} of inflation is a stochastic process, inflation will probably be eternal.  This is because if we consider a relatively large inflating region, we already know (from the double-well case) that if inflation ends by local and uncorrelated events, then even if the resulting non-inflating regions expand at lightspeed into the surrounding inflating region, they cannot eat up the entire inflating region in terms of its physical volume.
Moreover, if the potential is complicated, it might be considered unnatural to assume that the universe begins uniformly, as this would require just the sort of conspiracy in initial conditions that inflation was designed to help avoid.  Without such a conspiracy it is hard to see how inflation could end everywhere at once.

\subsection{Criticism of eternal inflation}
\label{sec-criticism}

While most inflationary theorists accept -- either embracingly or grudgingly -- the idea and importance of eternal inflation, it does have its critics.  The concerns of which I am aware might be crudely divided into `technical' objections about the physics details, and `philosophical' or perhaps `conceptual' concerns about how it is understood.

First, understanding stochastic eternal inflation correctly requires careful understanding of the back-reaction of quantum fluctuations on the spacetime, and it is unclear that the toy calculations in the literature are reliable.  For example~\cite{Mukhanov:1996ak} argues that back reaction becomes important well before the `stochastic' regime, so that the latter is not under good calculational control.  A similar argument was put forth by~\cite{MersiniHoughton:2007hb}. As Neil Turok bluntly puts it: `The calculations so far presented to justify eternal inflation in fact break every known principle in theoretical physics: they are neither properly quantum mechanical, nor coordinate invariant, hence they violate unitarity and energy-momentum conservation, Bianchi identities, and so on.'~\cite{Turok:2002yq}

However, I do not think these criticisms apply to eternal inflation in general, but really only to the specific case of stochastic eternal inflation. Eternality in topological inflation seems clear even at the purely classical level. The picture of false-vacuum eternal inflation on which I have focused here also seems fairly unambiguous.  The background space is de Sitter, which is well understood (at the classical level at least), and the Coleman-DeLuccia bubble nucleation process is as well-defined as anything else in semi-classical quantum gravity and does not appear to be controversial.  Given these ingredients, the structure of the eternally inflating mixture of bubbles and background space is well-defined {\em up to} considerations involving bubble collisions (see, e.g.~\cite{Aguirre:2007an,Garriga:2006hw,Hawking:1982ga}), and nucleations `up' the hill from the true vacuum to the false vacuum (see Sec.~\ref{sec-startinginf} below); both of these are much more poorly understood, but neither impacts the eternality of inflation. Moreover, I find it hard to believe that stochastic eternal inflation is somehow fundamentally ill-defined even while false-vacuum eternal is inflation all right, because both have essentially the same character, where pockets of non-inflation forms stochastically within a sea of inflation. 

Another criticism of eternal inflation is that while it may occur {\em in some manner of description}, it does not matter, because one should only ever consider the spacetime region causally accessible by a single worldline.  In some cases this view is motivated by the fact that events behind the horizon of an observer simply are not relevant for that observer's observations, thus physics should admit a complete description in terms that neglect that exterior.  Other argue more provocatively on grounds of `holography' that those outside regions really are just complentary descriptions of the regions within and on the horizon (e.g.,\cite{Goheer:2002vf,Banks:2002nm,Bousso:2006ge,Freivogel:2006xu}).

While these are interesting and in some cases compelling views, I would argue that they do not change the need understand eternal inflation in the usual sense of many horizon Hubble volumes.  For example, even if there exists a description of everything I will experience through the next 10 minutes in terms of quantities definable only within a 5 light-minute sphere, understanding the sun -- which is deeply important for comprehending many aspects of those 10 minutes -- in terms of these quantities would be extremely complex and unenlightening.

\section{Observables and probabilities in eternal inflation}
\label{sec-obsprobs}

Although inflation has some impressive observational successes, and in many cases leads naturally to eternal inflation, eternal inflation itself creates some profound difficulties regarding the link between the fundamental theory and our current and future observations and experiments.  The reason is that all of the pocket universe are equivalent only in the simplest models.  For more general potentials, inflationary predictions can differ from pocket to pocket (e.g.,~\cite{Aguirre:2006na,Tegmark:2004qd}).  Even more problematicly, the way in which high-energy symmetries are broken -- so as to determine low-energy effective physics -- could vary from pocket to pocket.  As an extreme and centrally important example, it is generally agreed that in string theory the compactification of extra dimensions leads to a complicated, many-dimensional effective potential `landscape' with different minima/vacua (e.g.,~\cite{Douglas:2003um,Susskind:2003kw}) corresponding to different observables within pocket universes created in that vacuum.  Thus there is no unique answer to `what should we observe in our bubble?'  While not the focus of the present article, it is worth briefly addressing these difficulties in the context of two related questions:

\begin{enumerate}
\item Is an eternally-inflating `multiverse' of the above-described type observationally distinguishable from a model with a finite period of inflation and a single realization of one particular type of low-energy physics and cosmology?

\item If there is a multiverse, many sets of low-energy physics and cosmology hold in different regions. How do we extract predictions for what we should observer in {\em ours}?

\end{enumerate}

\subsection{Can we observe other pocket universes?}

For some time it has been generally agreed that the answer to the first question is `no', because each pocket universe is spatially infinite and (as is clear from Fig.~\ref{fig-bubble} for example), one would have to travel faster than light or backward in time to escape one's pocket. Recently, however, it has been realized that because bubbles collide (e.g.\cite{Guth:1981uk}), and many observers within a bubble would have such collisions within their past lightcones~\cite{Garriga:2006hw}, that these collisions might actually leave an observable imprint on our bubble~\cite{Aguirre:2007an}.  

In particular, the study of~\cite{Aguirre:2007an} found that given the expected small nucleation rate of bubbles, there are two circumstances in which an observer might have bubble collisions to their past.  The first is at very late times, in which case the bubbles would be quite small and probably impact the observer's bubble only very slightly.  The second case is at early times for observers far from the `center'\footnote{An isolated  bubble nucleated in pure dS is homogeneous to observers inside it, and thus has no center.  But the bubble {\em distribution} in the exterior spacetime does carry a preferred frame in which it is statistically homogeneous (e.g.,\cite{Aguirre:2003ck}), and this defines a center of a bubble in terms of the impacts of other bubbles~\cite{Garriga:2006hw}.} of the bubble (and in an infinite bubble essentially all observers would be far from this center).  In this case the impact can be severe and affect essentially all of the observer's sky, and the question is whether the collision would preclude the observer's very existence.  Work on resolving this issue, and in specifying in greater detail the potential observational signatures in, say, the Cosmic Microwave Background (CMB) is ongoing. (One thing that seems concretely predictable is that the signature of large bubbles would be azimuthally symmetric about some particular direction in the sky; one could therefore search for such a signal, which is not predicted by the standard inflationary perturbation mechanism.) At minimum, this work is proof-of-principle that the inflationary multiverse may lead to directly observable signatures.\footnote{Open inflation itself carries possible observational signatures including nonzero curvature ($\Omega \neq 1$) and large-scale signatures in the CMB (see, e.g.,~\cite{Garriga:1997ht,Garriga:1998he}).  Some of these would already constitute evidence for a multiverse if open inflation could only arise via decay from a valse vacuum; but this is not clear: see~\cite{1998PhLB..425...25H}. Whether there are signatures other than bubble collisions that indicate a parent false-vacuum, along the lines of~\cite{Garriga:1997ht}, is an interesting and pertinent question.}

\subsection{Typicality and the measure problem}

While observing effects other pocket universes would be extremely interesting, and directly bolster the eternally-inflating model, it is rather unclear that it would resolve the second question, of how to compare our universe to a fundamental theory predicting many such regions with diverse properties.

When this problem is first considered it does not seem so difficult: why not just calculate which universe types are common, and which rare, and assume that we should be in one of the common ones?  But be assured, the more you think about the issue the more terrible it will become.

This terribleness has two aspects.  The first is a philosophical, or perhaps methodological, problem:  what do we {\em assume} in making our calculation?  One way to pose the issue to is form questions such as `given that our observations are like those of a randomly chosen X, what should we observer?', where `X' might stand for universe, or point in space, or observer, etc.~\cite{Aguirre:2005cj,Aguirre:2004qb,bostrombook}. Two things are immediately clear.  First, the answer to the question will almost certainly depend on what X is, and second, it is rather unclear, even in principle, to which sort of X our observations should be most similar.  

One might hope to evade these difficulties by re-posing the question.  For example, we might strike `randomly chosen', but this accomplishes nothing since absent any further information we could only choose our X at random.  Or we could try to `just look for correlations between observables'.  This is fine, but insufficient, as the correlations will certainly be statistical, with the statistics depending, again, on what conditioning one does.  Or we could hope to replicate what we do in the lab: narrow our focus only to regions with identical properties to ours in terms of what has been measured so far, and make predictions for the future.  But not only does this render all of our current knowledge nearly useless in terms of checking our theory, but it also will favor or rule out different theories depending upon when the testing is done: what is ruled out on the basis of an experiment done today could be consistently accepted tomorrow as `input data'.  These thorny issues are discussed at length in the literature; see, for example~\cite{bostrombook,carrunivermulti}.

The second terrible things is that {\em even if we decide} on a particular X, there are still fundamental ambiguities in assigning probabilities to this X (essentially via counting Xs) in eternal inflation.  There are many potential ways to do it, with reasonable motivation, but with different results; see recent reviews by~\cite{Aguirre:2006ak,Vanchurin:2006qp,Winitzki:2006rn,Vilenkin:2006xv,Linde:2006nw}.  The root of the ambiguity seems to be that (a) the quantities being compared are in all cases infinite, and (b) that the exponential expansion and lack of clearly-useful symmetry in eternally inflating spacetimes prevents the choice of a unique, `obvious' regularization of these infinities.  It seems likely that further work on this problem will lead to further progress and insight (indeed the recent spate of work cited above has brought a great deal more order to the subject), but at the moment the problem still seems rather open.

\section{Inflation from non-inflation}
\label{sec-startinginf}

As discussed in Chapter 2, in inflationary theory the creation of a \hotpocket\ occurs when inflation ends; and we have seen that in eternal inflation the link between the well-tested `big bang theory' of a hot dense, evolving space, and the `big-bang' as the beginning of time is nearly severed: the inflating phase can have lasted for an indefinitely long time before reheating.

But we can still pose the question:  how did inflation start?  It would seem there are three basic (though as we shall somewhat overlapping) options:

\begin{enumerate}

\item At  `the beginning of time' inflation sprang from a primordial singularity or some other system that cannot be described using classical general relativity (GR) -- i.e. in much the same way posited by the classic big-bang theory.

\item Inflation sprang somehow from a non-inflating region of classically well-defined spacetime.

\item The universe was always inflating.

\end{enumerate}

The first option is discussed in various ways in many chapters of this volume, so I focus here on the second (this section) and third (rest of the chapter) possibilities.

\subsection{Classical Casuality and Inflation}

The question of whether an inflationary universe could be created simply by forming a small region of high vacuum energy was first addressed in detail by Farhi \& Guth~\cite{Farhi:1986ty}, who asked: given sufficient technology, can I create a large enough region that it will begin inflating, and thus create a whole universe (or more)?

To the disappointment of aspiring Creators everywhere, they discovered that there is a fundamental obstacle: a small region of false-vacuum has energy, and before you can make a false-vacuum region large enough to inflate, it inevitably collapses into a black hole.  Let's look at this statement and its subtle implications in a more detailed way; we will get some good news, then some bad news, then good, then bad, then some more bad, then a ray of hope.  

\begin{figure}
\centering
\includegraphics[width=12cm]{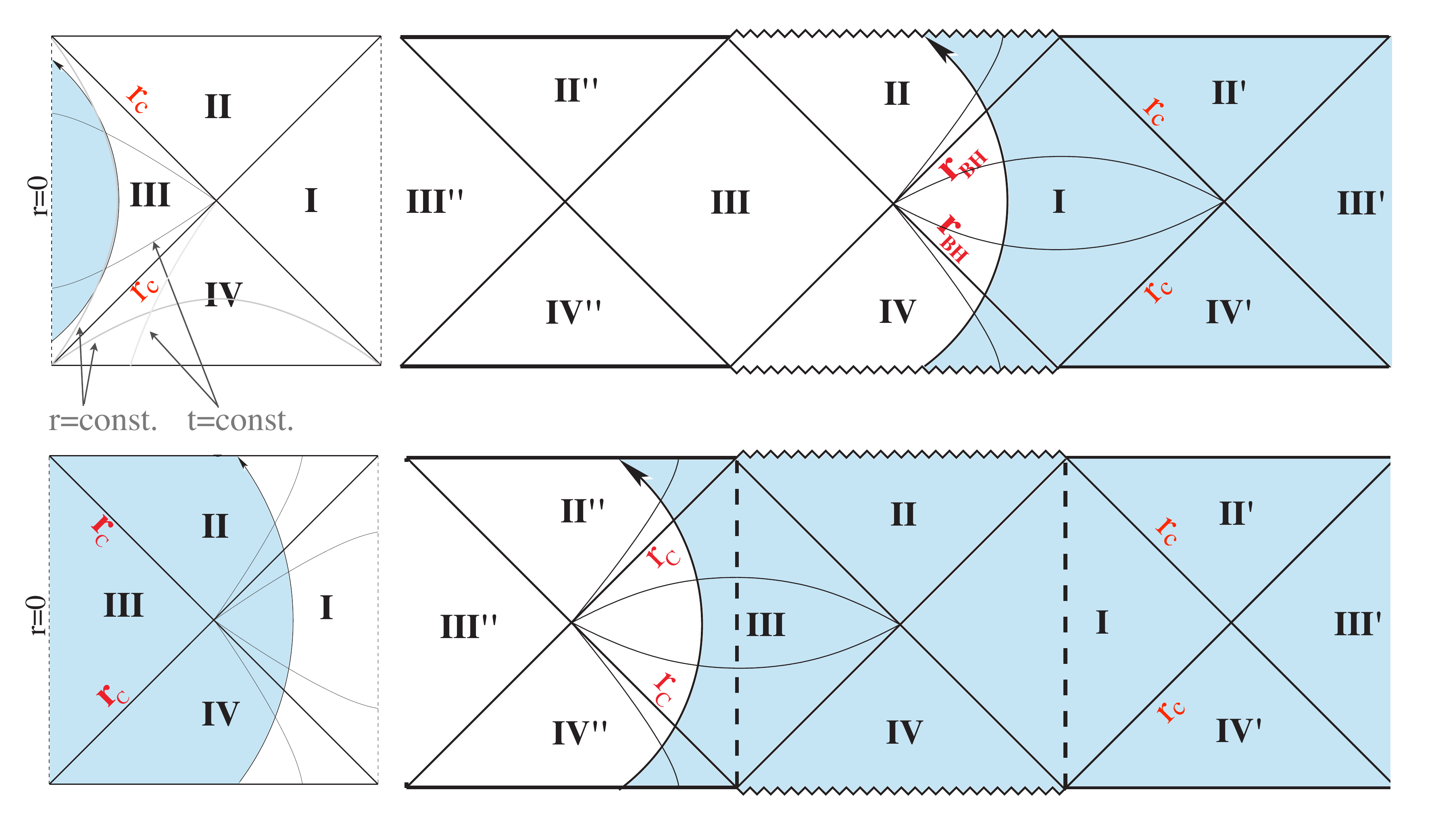}
\caption{Exact solutions representing an inflating region within a non-inflating background, obtained by gluing de Sitter (dS, left) to Schwarzschild-dS (SdS, right) along a domain wall.
Both are conformal diagrams (radial null paths are diagonal, each point represents a 2-sphere) of the maximally extended spacetimes.  In dS, patch I or patch III is covered by static coordinates giving metric~\ref{eq-sdsmetric} with $M=0$. Several surfaces of constant $r$ and $t$ are shown; other patches are covered by analytic continuations of these coordinates.  The null surface $r_{c}$ is the cosmological horizon of an $r=0$ observer.  In SdS, an observer in region I sees a cosmological horizon at $r_c$ (between regions I and II'), a black hole horizon between I and II at $r_{\rm BH}$, and a `white hole' naked singularity in region IV.  The (spherical) wall is indicated by the arrow and separates the physical (shaded) dS region from the physical (shaded) SdS region.
\label{fig-gluing}}
\end{figure}

A good way to get a firm handle on the issue is by examining exact solutions to Einstein's equations that describe an inflationary region embedded in a non-inflationary one.  This is tractible under the assumptions that both spacetimes are spherically symmetric, and that they are `glued' together by an infinitely thin `wall' of some fixed tension (i.e. fixed energy per unit area in its rest frame).  With these assumptions, both interior and exterior spacetimes must be (if for simplicity we also assume charge neutrality) Schwarzschild-de Sitter (SdS), which looks static in the coordinates giving metric:
\begin{equation}
ds^2=-f(r; M, \Lambda)c^2dt^2+f^{-1}(r; M,\Lambda)dr^2+r^2d\Omega^2,
\label{eq-sdsmetric}
\end{equation}
where $f=(1-2GM/c^2 r-\Lambda r^2/3c^2)$; this essentially describes a black hole of mass $M$ embedded in dS with cosmological constant $\Lambda$ and thus vacuum energy density $c^4\Lambda/8\pi G$. We can then glue a spherical bubble of inflating dS ($M=0$, $\Lambda$ large) into a background dS or Minkowski spacetime ($M$ arbitrary, $\Lambda$ small or zero).  The construction is as shown in Fig.~\ref{fig-gluing}.  Starting at the top-left, we have a conformal diagram for dS, where the diamond labeled `III' is covered by $0 \le r < \infty$, $-\infty < t < \infty$; the other regions can be covered by other coordinate patches.  The diagonal lines signify dS horizons.  On the top-right, we have maximally-extended $M\neq 0$ SdS, where an observer in region I sees a black-hole horizon as labeled $r_{\rm BH}$, and a cosmological horizon at the boundary of regions I and II'.  Note that the spacetime is {\em periodic}: region II'' is just like II'; regions III'' is just like I; region III' is just like III, etc.\footnote{Note also that region IV is a `white hole', which is the time-reverse of a black-hole: it has a past singularity and a region (IV) that one can leave but never enter.} Now, let the bubble wall be indicated by the curved arrow; the shaded region to the right is physical, while the white area is replaced by the shaded area on the dS diagram to the left, on the other side of the wall.  The dynamics of the wall are determined by the matching conditions across the wall that Einstein's equations require; see, e.g.~\cite{Aguirre:2005sv,Aurilia:1989sb,Berezin:1982ur,Blau:1986cw,Cvetic:1992jk}

Voila! Good news! We have an high-vacuum-energy region embedded in a non-inflating region.  Moreover, although in this model the bubble emanates from a past singularity, we can {\em imagine} replacing the bottom half of the diagram by one in which there is no bubble, i.e. in which we create the bubble at some time out of some other raw materials. But alas (bad news), examination of the dS diagram shows that rather than creating a universe, the bubble last for just a short time before collapsing to zero radius (From the outside, it looks like a bubble that shoots out of a white-hole, then recollapses into a black hole.)  Undeterred, we make the bubble bigger. And eventually, this works (good news): we get the diagram on the bottom-left, in which the bubble wall expands forever (to both the past and to the future).  But looking at the bottom-right diagram, we see (bad news) that the whole bubble trajectory is behind the horizon. But then {\em how do we create the bubble?}  This is the nub of the problem.

The next piece of bad news is that this problem is quite general, as shown by Guth and Farhi~\cite{Farhi:1986ty}.  In particular, they assumed only a rather weak condition on the allowed form of the energy-momentum tensor (the `dominant energy condition'; see Appendix), and used a singularity theorem due to Penrose~\cite{Penrose:1964wq} (see also Sec.~\ref{sec-singth} below)  to show that any seed capable of inflating to the future would inevitably have a singularity to its past.  Thus to create an inflating universe we would {\em either} have to violate this energy condition, or create (or find, I suppose) a naked singularity.  Both are extremely unappealing possibilities.

A second, and similarly discouraging, angle on this was provided in~\cite{Vachaspati:1998dy,Dutta:2005gt}, where it was shown that if the `null energy condition' (see Appendix) holds, then an for an inflating region to exist with a size greater than its inflationary horizon size, it must take up a region larger than the horizon size of the exterior spacetime in which it is embedded.  
This would require a region several Gigaparsecs in size for our current universe, making the task rather difficult. Admittedly, the key tool in showing this is the ability to follow a set of null geodesics through the boundary between the interior and exterior regions, therefore if an obstruction such as a singularity exists the basic idea can be circumvented (as in the bottom solution shown in the Fig.~\ref{fig-gluing}); allowing topology change can also help circumvent the idea~\cite{Borde:1998wa}, and there are other subtleties~\cite{Aguirre:2005sv}.  Nonetheless the result, like that of Guth and Farhi, points to a basic reluctance by nature to create inflating regions within non-inflating ones in a causal way.

The key import of these question is for the age-old question: `is there such thing as a free lunch'?  It is often claimed that inflation is the {\em ultimate} free lunch.  Indeed, if you want to create a large or even infinite universe (or infinite set of them) inflation will do it.  But to get inflation itself, it seems that you need to either (a) break causality (in the form of violating energy conditions), or (b) create/use a naked singularity (violating cosmic censorship), or (c) turn a region larger than the cosmological horizon into an inflating region.  

Thus inflation is more like an endless, all-you-can eat buffet, but with an admission price that is unaffordably high.  It's so expensive that the only way to pay is to win the lottery.  And so we turn to quantum mechanics.

\subsection{Quantum nucleation of inflating regions}

Stymied in their theoretical efforts to produce a universe in the laboratory, Farhi, Guth \& Guven~\cite{Farhi:1989yr}, as well as Fischler, Morgan \& Polchinski~\cite{Fischler:1989se,Fischler:1990pk} and later Linde~\cite{Linde:1991sk} investigated whether a quantum tunneling process might allow the creation of a so-called `baby universe'.

\begin{figure}
\centering
\includegraphics[width=0.5\textwidth]{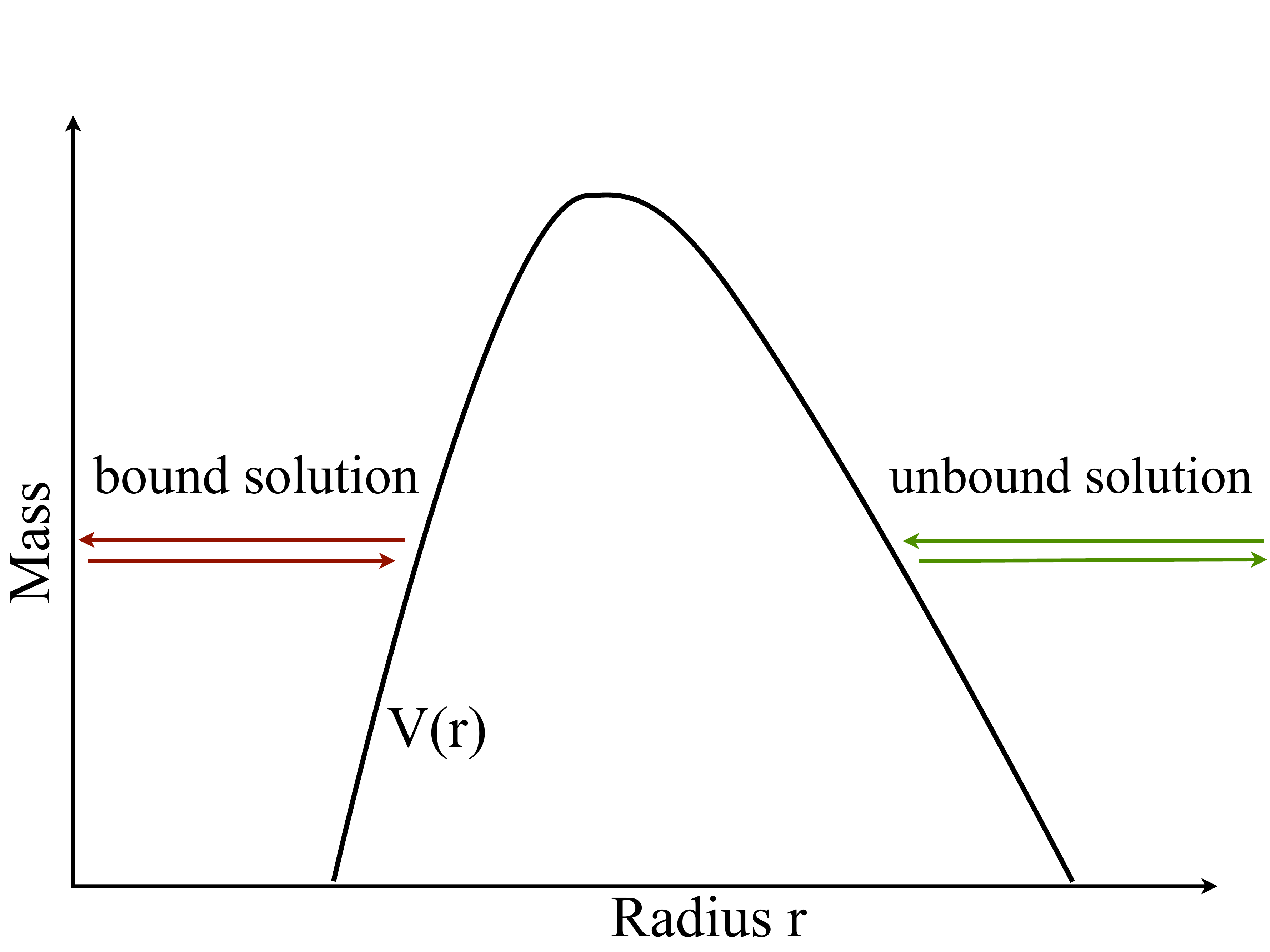}\includegraphics[width=0.5\textwidth]{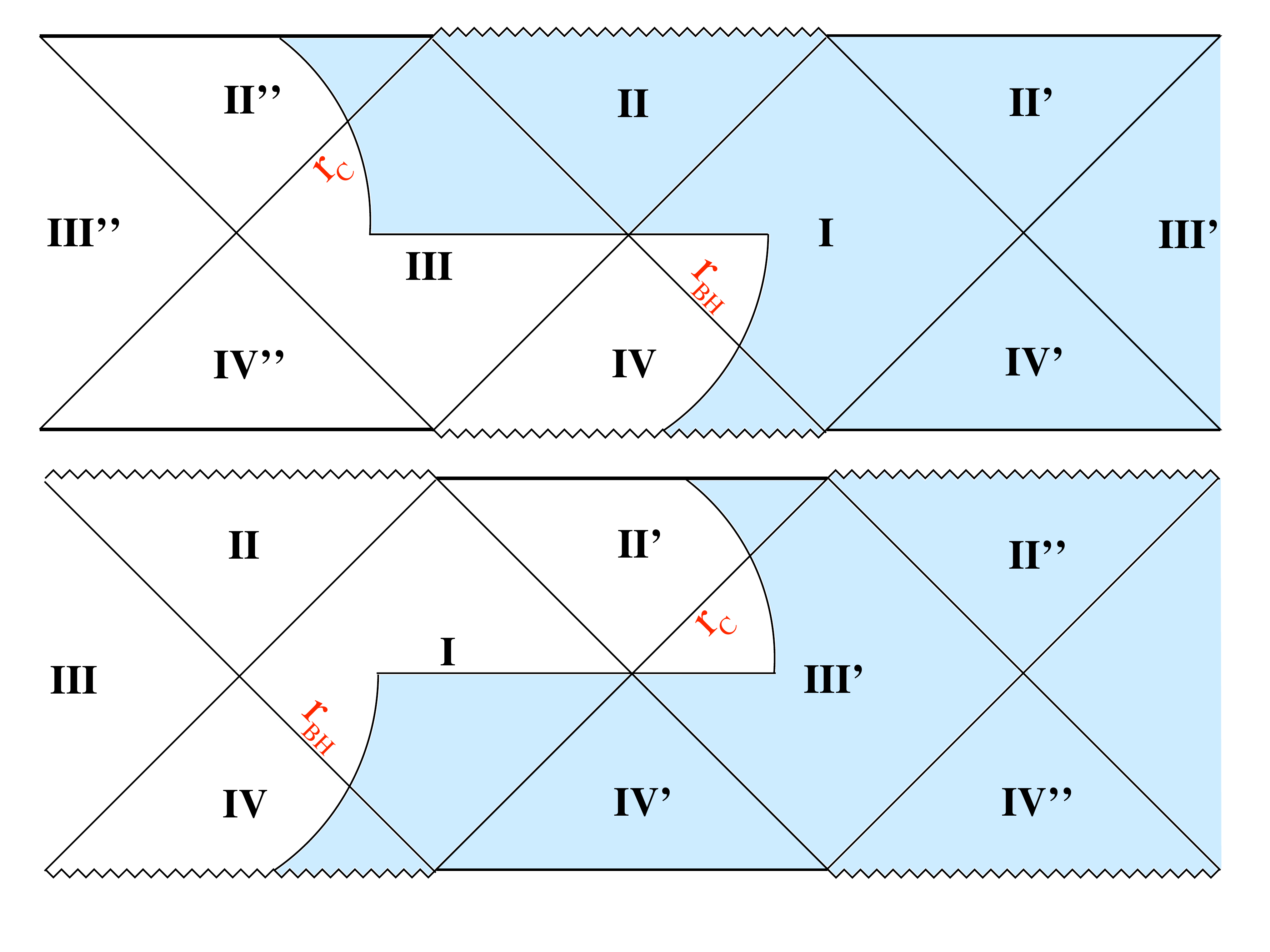} \hfill
\caption{{\em Left:} Bubble-wall dynamics as a 1-D particle-in-potential problem. The two spacetimes sketch in Fig.~\ref{fig-gluing} correspond to the `bound' and `unbound' solutions shown. {\em Right:} Top: `acausal' up-tunneling as described by~\cite{Fischler:1990pk,Farhi:1989yr} in which a small bubble in region I tunnels through the black-hole horizon to become a large bubble in region III.  Bottom: `causal' up-tunneling (a generalization of the process described by~\cite{Lee:1987qc}) in which a small bubble tunnels through the cosmological horizon to become a large bubble in region III'.
\label{fig-tunneling}}
\end{figure}

The appeal of this idea is that the equations of motion for the bubble wall radius $r$ can be recast into the very simple form of a 1D particle in a potential $V(r)$ (see, e.g.,\cite{Fischler:1990pk,Farhi:1989yr,Blau:1986cw,Aurilia:1989sb,Aguirre:2005cj,Aguirre:2005nt}).  The energy of the `particle' is given by the bubble mass; the potential depends (but only weakly) on the bubble wall tension and inner and outer vacuum energies; see Fig.~\ref{fig-tunneling} for a sketch.

Each `glued' spacetime corresponds to a trajectory on this potential diagram; for example the top and bottom spacetimes in Fig.~\ref{fig-gluing} correspond, respectively, to the  `bound solution'  and `unbound solution' of Fig.~\ref{fig-tunneling}. In fact, earlier I elided over the fact that these can be of the {\em same mass}: we can make the bubble bigger in radius by increasing $M$ for a while, but after that, the wall disappears behind the black hole horizon; thereafter, we can get a yet bigger bubble by {\em decreasing} $M$. 
We can thus effectively quantize our 1D particle and imagine a fixed-energy tunneling process between the bound `seed' (which we can create in a causal way) into the unbound inflating bubble (which we cannot create ourselves).  This is exactly the process described by~\cite{Fischler:1990pk,Farhi:1989yr}, and it looks like the top-right panel of Fig.~\ref{fig-tunneling}.  This idea is very nice, but has some troubling qualities.  For one, it is still {\em acausal}: the post-tunneling bubble is completely disconnected from the original spacetime; once it forms no observer can ever go into it, or even see it.  Second, in the $M\rightarrow 0$ limit it becomes a bit strange, in that the original spacetime looks like just unperturbed dS or Minkowski space and appears to play no role whatsoever in the creation of the baby universe. There are also troubling technical issues (see~\cite{Aguirre:2005nt,Ansoldi:2007it} and references therein for extensive but rather inconclusive discussion).  

Oddly, though, there has long been {\em another} description of tunneling into a false vacuum.  Lee \& Wienberg~\cite{Lee:1987qc} noticed that the same instanton found by CDL that mediated decay of the true vacuum {\em also} described the embedding of a false-vacuum bubble in a true-vacuum exterior.  The instanton could then be considered to describe a process in which an absolutely enormous true-vacuum region (larger than the true-vacuum horizon size) spontaneously `hops' into the false vacuum (see also~\cite{Garriga:1993fh,Garriga:1997ef}).  The probability for this to occur is absurdly small, as it can be interpreted as a downward fluctuation in entropy (to essentially zero) of the entire Hubble volume.  This probability much smaller even than the already-tiny chance of the `acausal' process described above.  
But in contrast, this process is causal: once the bubble forms, one could, in principle, jump inside.
		
These two processes were sufficiently different from each other, individually complex, and stupendously improbable (hence largely ignorable), that their relation to each other remained obscure for quite some time.  But as it turns out, they are nearly the same!  More specifically, they decribe tunneling {\em from} same spacetime, and {\em to} the same spacetime, on the same potential diagram, but via a different path~\cite{Aguirre:2005nt}.  In particular, a slight generalization of the `causal' Lee-Weinberg process is as a small seed that tunnels into a large bubble, as per the bottom panel of Fig.~\ref{fig-tunneling}. Now recall that the basic spacetime, SdS, is periodic.  Thus if we slice both of Fig.~\ref{fig-tunneling}'s conformal diagrams horizontally through the middle, we are free to `translate' the upper and lower half without doing anything at all: IV' (say) becomes IV'', or vice-versa, etc.; the only difference is which part we shade.  Now we can see that if we take the top panel (acausal tunneling) and shift the top half to the right, and the bottom half to the left, we obtain the bottom panel (causal tunneling).  This does not mean the processes are the {\em same}, but clearly they are closely related.

Which process actually occurs?  This is unknown: both involve both quantum mechanics and relativistic gravity inextricably, so we are on very shaky ground.  It may take a true theory of quantum gravity to clarify the issue (string theory, incidentally, so far is of no help whatever).  The causal process seems less controversial,
 but this does not make the question moot, because the acausal mechanism, if it occurs, would be vastly more probable.  And as discussed next, the two would lead to rather different pictures of the large-scale structure of a universe in which they occur.

\subsection{Recycling in eternal inflation}
\label{sec-recycle}

A picture of eternal inflation including causal up-jumps was put forward by~\cite{Garriga:1997ef} who called it the `recycling universe.'  A simple model is given by the double-well potential of Fig.~\ref{fig-doublewell}, with $V(\phi_T) > 0$. Within each bubble of true vacuum, large bubbles of false-vacuum eventually form; within each of these, infinitely many new true-vacuum bubbles, and so on.  In this view, any given worldline (and, neglecting the relatively brief transition period, any given Hubble volume) endlessly cycles between true and false vacuum.  Stochastic eternal inflation, which also supports upward jumps whenever the potential is positive, could be considered a special case of this scenario.

As pointed out by~\cite{Dyson:2002pf}, this picture runs into trouble {\em if} one takes the view -- mentioned in Sec.~\ref{sec-criticism} -- that only one Hubble-volume surrounding a single worldline should ever be discussed, and also that a recycling event lies to our past.  Consider some set of observational data to be explained.  In this model upward jumps, followed by true-vacuum bubble nucleations, would indeed create \hotpocket s that could explain the data.  But there would also be freakish downward entropy fluctuations from equilibrium that would also reproduce that observational data.  While seemingly-miraculous, these would nonetheless be exponentially more common than the upward (in vacuum energy) transition to inflation.  Because the properties of these `freak' regions would diverge from the more `natural' evolution when concerning any observation not yet made, we can quickly rule out this scenario.\footnote{It seems clear that this is a fatal problem for any model of a finite system that {\em appeals} to a downward entropy fluctuation to supply an `initial' (low entropy) state.  One might further argue -- but I think less convincingly -- that any time a system creates `freak observers' (even in as simple form as an insolated brain) that are more abundant than `natural' ones, there is a similar problem.}

There are a number of way in which this problem may be avoided.  The basic ideas of which I am aware are:

\begin{enumerate}

\item One might appeal to inflation to transform a small volume that has fluctuated far down in entropy {\em density} and turn it into a large region of low entropy density~\cite{2004PhRvD..70f3528A}.  But this is exactly what inflation refuses to do in a causal way. That is, the new inflating region, even if small, is either `expensive' as an entropy fluctuation, or is behind a wormhole so that the connection to a pre-existing worldline is largely cut.  Still, the acausal up-jump mechanism, if it occurs, would indeed create many false-vaccum regions and might solve the puzzle.

\item{If the global potential minimum is negative, then eventually any position in space will tunnel to a region with negative vacuum energy and end in a big-crunch singularity.  If this almost always occurs before a freak region is created, and the singularity is taken to be a true `end of time' from which nothing can ever emerge, then the problem might be solved (e.g.,~\cite{Bousso:2006xc,Page:2006nt})}
	
\item One might argue that the notion of a fixed set of states, each of which gets ergodically sampled in the infinite-time limit, is inapplicable to cosmology, or otherwise argue that the {\em first} realization of a \hotpocket\ is sufficient and/or categorically different from subsequent realizations and should not be compared or lumped with them (e.g.,~\cite{Banks:2007ei}; see also~\cite{Hartle:2007zv}).
	
\item If the restriction to a single Hubble volume is removed, then one might argue that in the full spacetime, there are infinitely many freak regions and natural regions, but that when you {\em count them} (see Sec.~\ref{sec-obsprobs}) the natural regions greatly outnumber the freaks~\cite{Linde:2006nw,Vilenkin:2006qg}.

\end{enumerate}

All of these seem to have the character of either (a) having an infinite number of states, so that entropy can continually increase without an equilibrium being reached, or (b) having a set of states that evolves, to either avoid equilibrium or to disconnect the equilibrated region from our observable universe. That is, the case seems quite tight against the universe being in some sort of equilibrium state.  Does this mean that by `pure thought' we have proven that a universe described by a finite number of states -- {\em no matter how large} -- does not make sense~\cite{aguirreinfinity}?  This is a remarkable thought that I shall leave here, noting only that this issue runs straight into three deep mysteries: how we place a measure on an eternally-inflating spacetime, how and whether baby universes form (i.e. causally or acausally, etc.), and what happens `after' a big-crunch cosmological singularity.

\section{Must inflation have a beginning? Singularity theorems and inflation}
\label{sec-singth}

Underlying the widespread belief that the universe began at some definite `initial time' (before which classical GR breaks down) is a product of two factors.  First is the wide array of convincing evidence that our observable universe evolved from a \hotpocket\ of $\gsim \,$MeV temperature.  The second is a set of results in classical GR showing that under (what seem like) rather general conditions, an expanding, radiation-dominated universe is singular to the past.  We have already seen that inflation changes the implication of the first factor, since a universe just like ours could have emerged from a long and complex inflating phase.  What about the singularity theorems?

Four key classic singularity theorems are treated, for example, in Wald~\cite{1984gere.book.....W}.  Three of them (theorems 9.5.1, 9.5.2, and 9.5.4) require the strong energy condition (see Appendix). This condition is violated by dS, and in general for spacetimes with accelerated expansion, so these theorems are simply not applicable to inflationary spacetimes.  The remaining one, due to Penrose~\cite{Penrose:1964wq} and mentioned above, holds using only the weak energy condition (see Appendix), and states that if a spacetime contains both (a) an anti-trapped surface\footnote{Roughly, an anti-trapped surface is one for which a family of null geodesics converges after emanating perpendicularly from {\em either side} of the surface.  Thinking of a spherical surface will quickly give you a sense of how weird this is.} and (b) a {\em non-compact} Cauchy surface\footnote{Roughly, a Cauchy surface ${\cal C}$ for a spacetime ${\cal S}$ is one for which (a) no two points are connectible by a non-spacelike path, and (b) for any point in ${\cal S}$, all past-directed (or alternatively all future-directed) non-spacelike paths intersect ${\cal C}$.}, then it has at least one past-directed geodesic of finite length. This geodesic that `ends' is taken to point to a singularity.

Inflationary spacetimes often contain anti-trapped surfaces of the required type, but this does not necessarily render them singular by the theorem.  To see this, and to gain insight into the import of other singularity theorems, it is very useful to consider dS, which is both (a) geodesically complete, and (b) the basic structure underlying both inflation and eternal inflation.  Any singularity theorem must therefore either fail to apply to dS, or only apply to {\em part} of it.

Take, for example, Penrose's result as applied to dS in static coordinates (Fig.~\ref{fig-gluing}).  Any point on the diagram in region II represents an anti-trapped two-sphere, and moreover any constant-$r$ surface is a non-compact Cauchy surface {\em for region II}.  The theorem thus applies to region II, and implies that it contains incomplete geodesics.  But this simply indicates that region II is extendible (into full dS).  And full dS admits only compact Cauchy surfaces so that the theorem does not apply to it.
Now this does not mean the theorem is useless -- indeed In Sec.~\ref{sec-startinginf} we saw that it provided great insight into the creation of inflation from non-inflationary ones. But this analysis {\em does} mean that we must interpret this theorem and others with great care, and suggests that it may not indicate singularities in a spacetime that is essentially dS.

\subsection{Inflationary singularity theorems}

The realization that the classic singularity theorems do not really address inflationary spacetimes led researchers to develop theorems that do; a nice review of these results is given in~\cite{Borde:1996pt}.  These theorems share some similarities with the classic theorems in, e.g., assuming both energy conditions and some properties of the global spacetime.  They also assume some conditions that are interpreted as indicating that the spacetime is inflating or eternally inflating.  These results are, I think, quite interesting; but even their authors did not consider them the last word, because many versions of eternal inflation violate even the rather weak energy conditions assumed by the theorems~\cite{Borde:1997pp}. This motivated Borde, Guth, and Vilenkin~\cite{Borde:2001nh} (hereafter BGV) to look for a result that did not require {\em any} energy condition -- and they found one!  Because this result is widely cited as definitive evidence that inflation cannot be eternal to the past, it is worth examining this theorem in some detail, especially because -- as explained below -- I think that this inference is incorrect.

Imagine that we have a spacetime that is expanding.  What does that mean?  A standard way to describe this is to define a `congruence of geodesics' (roughly, a dense set of geodesics that do not cross); each geodesic would be the worldline of a test particle with coordinates $x^\mu(\tau)$, where $\tau$ is the proper time experienced by the imagined particle.  The `expansion'\footnote{Note that this is {\em not} the  `expansion' in a comoving congruence that is often used in singularity theorems and is governed by Raychaduri's equation; see, e.g.~\cite{1975lsss.book.....H}.} of this set of particles can then be discussed in terms of the vector field $u^\mu\equiv dx^{\mu}(\tau)/d\tau$: roughly, if these 4-velocity vector `point away from each other' in space with increasing time, then (that region of) the universe is expanding.  

A precise sense of this was defined by BGV, as follows.  Define an equal-time surface using the proper time of the geodesics in some region.  Then, at a single time, we can define a (spacelike) vector connecting two neighboring geodesics in the congruence (call it $\delta r^\mu$), as well as a spatial separation vector (call it $\delta \vec r$), and the radial part of the two geodesics' relative velocities (call it $\delta u_r$).  A `Hubble constant' can then be defined locally by 
$
H\equiv {\delta u_r / \delta r},
$
in the limit of small $\delta r\equiv |\delta\vec r|$.  BGV show that this quantity can be determined by an observer along a worldline just by measuring the local velocity of the geodesics in the congruence relative to the observer, and noting how they change as the observer moves along his/her worldline.  The definition is constructed to correspond to the usual definition $H=(1/a)da/dt$ of the Hubble parameter in an FLRW cosmology, and the method of measurement corresponds to the fact that in an expanding universe, peculiar velocities decay away. (Intuitively, this is because an observer passing some particle with some velocity will have a progressively harder time `catching up' to farther and farther particles that move faster and faster away from the first.) Given this definition, BGV show that following an observer into the past along its worldline, there is a bound on the integral of $H$ over the observer's proper time.  Therefore if $H$ always exceeds some positive number, then the past-proper time of the observer must be finite, hence the worldline geodesically incomplete.   

Does this result (hereafter denoted `\result') mean that past-eternal inflation is not possible?  BGV carefully phrase their conclusion as showing that there is a boundary to the inflating region, and supplying a few speculations as to what that boundary is.  But in the inflationary `lore' the interpretation of the theorem is that it rules out past-eternal inflation.  I strongly disagree with this interpretation; the next few sections will explain why.

A first hint of concern arises by noting that not {\em all} geodesics are incomplete: in particular, \result\  does not apply to the comoving geodesics themselves!  Thus there is nothing preventing some geodesics from continuing arbitrarily far into the past.  This is clearly worrisome; for example in {\em future} eternal inflation only a subset of geodesics of measure zero are inflating into the future with infinite proper time.

A second worry is that \result\  is so general that it does not depend on any energy conditions.  Why should this be worrisome?  Because {\em any} spacetime solves Einstein's equations $G_{\mu\nu}={8\pi G \over c^4} T_{\mu\nu}$ for some energy-momentum distribution $T_{\mu\nu}$: simply pick any desired metric, compute $G_{\mu\nu}$ from it, and divide by $8\pi G/c^4$!  Without any constraint on the forms that $T_{\mu\nu}$ can take, there is nothing to be said against the result. Thus the widespread interpretation of \result\ is essentially that a forever-expanding metric spacetime is simply inconceivable -- and this seems hard to accept.

A third concern, which seems rather obscure but will emerge as central, is that `expansion' versus `contraction' depends on the direction of time. That is, reversing time will convert an expanding region into a contracting one.  Note that this is {\em not} true, for example, for the {\em acceleration} of particles toward or away from each other.  

\begin{figure}
\centering
\includegraphics[width=12cm]{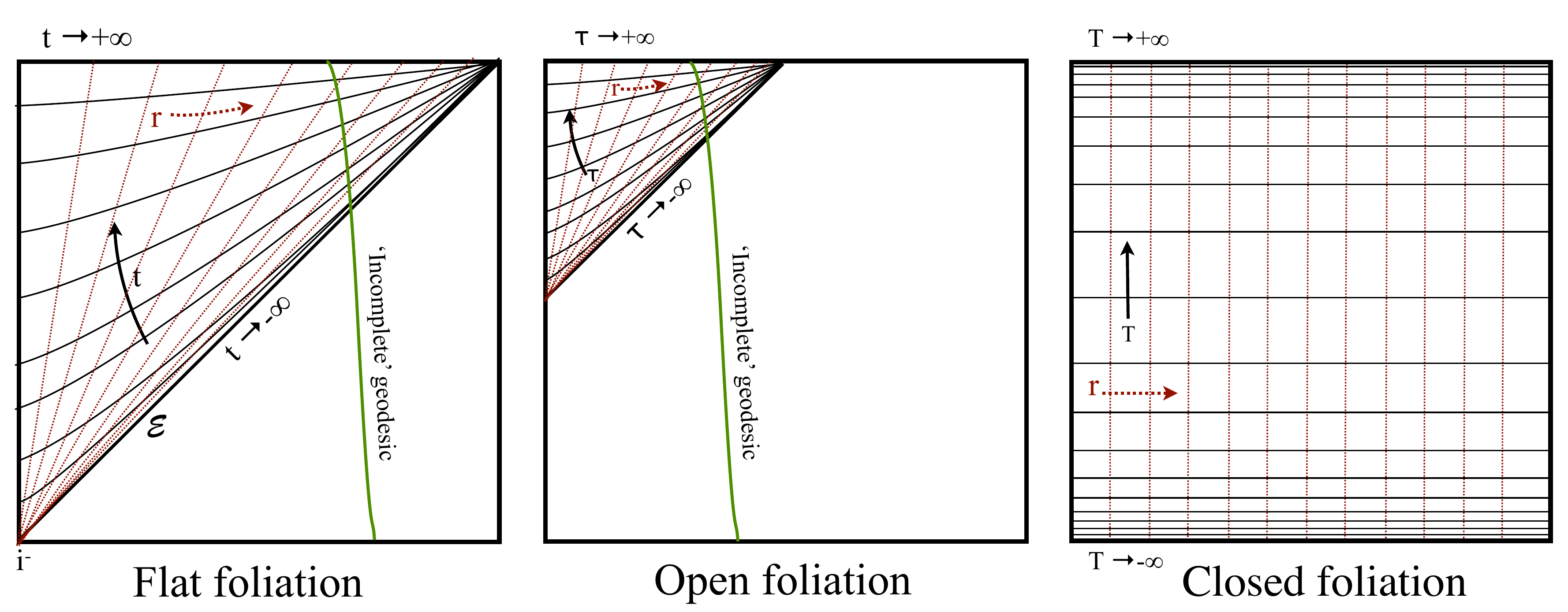}
\caption{De Sitter space (dS), the spacetime underlying inflation and eternal inflation. dS is a D-dimensional space of constant scalar curvature and is a solution to Einstein's equations for a positive cosmological constant and no other material contents.  It admits three foliations (here denoted `flat', `open' and `closed') into constant-time surfaces that are D-1-dimensional surfaces of constant curvature; in each of these the spacetime is then a FLRW model.  In the flat foliation (left panel) the metric can be written $ds^2=-c^2dt^2+\exp(2Ht)[dr^2+r^2d\Omega^2]$ with $-\infty < t < \infty$ and $0 < r < \infty$; but the coordinates cover only half of the spacetime and $t \rightarrow -\infty$ approaches a null surface \edge\ as shown.  In the open foliation, the metric can be written $ds^2=-c^2d\tau^2+\sinh^2(H\tau)[d\xi^2+\sinh^2(\xi)d\Omega^2]$ with $0 < \tau < \infty$ and $0 < \xi < \infty$; but the coordinates cover only the interior of a lightcone as shown. In the closed foliation, the metric is $ds^2=-c^2dT^2+\cosh^2(HT)[d\eta^2+\sin^2(\eta)d\Omega^2]$ with $-\infty < T < \infty$ and $0 < \eta < \pi$; these coordinates cover the full spacetime.  
dS also admits a static foliation as discussed in Fig.~\ref{fig-gluing}.
A timelike geodesic can leave either the flat or open charts in the direction of decreasing $t$ or $\tau$.
\label{fig-dsreview}}
\end{figure}

To build upon this concern, let us apply \result\ to dS, as we did with Penrose's result above.  As reviewed in Figure~\ref{fig-dsreview}, there are three congruences on dS that could describe a homogeneous distribution of particles; in each congruence a geodesic corresponds to a fixed coordinate position, and a surface of constant coordinate time is a uniform space of either positive, negative, or zero constant curvature.  

Applying \result\ to dS using each of these congruences shows us exactly why non-comoving geodesics with $H > 0$ have finite length. What happens in the flat and negatively curved foliations is that our geodesic reaches the edge of the region of dS covered by the coordinates.  We can see this easily in Fig.~\ref{fig-dsreview}: for the open and flat foliations, the covered region is bounded by a null surface to the past, and one can easily draw a wordline that encounters this boundary. 
Only worldlines that approach the comoving ones stay within the covered region far to the past -- but note that in the flat foliation those geodesics do indeed extend forever into the past with $H > 0$.

Now, does this mean dS is singular?  Certainly not: it means that the {\em region over which the assumptions of \result\ hold} has a boundary.  Thus it is very important to analyze carefully exactly what those assumptions mean, and what they don't.  In this case, they can clearly say little about the actual spacetime, since all points of dS are exactly equivalent -- so there is no sense in saying that part of it has certain properties (such as being inflating) and other parts do not. And beyond the `edge' to which the theorems point is just more of {\em exactly the same} spacetime.

Let's go on to examine the closed coordinates (see Fig.~\ref{fig-dsreview}), which cover all of dS.  There is nowhere for our geodesic to `escape to', so what does \result\ mean here?  Here we see that if we follow one of our geodesics back, at some point $H$ switches to being negative past this point: the congruence is `contracting'.  But as before dS does not expand in some places and contract in others -- all points are equivalent; moreover dS obeys time-reversal symmetry, yet time-reversal would convert an `expanding' region into a `contracting' one, and vice-versa. dS does {\em accelerate} neighboring geodesics away from each other (everywhere, and in both time-directions), and this is directly related (via Raychaduri's equation) to the fact that the vacuum energy sustaining dS violates the strong energy condition.  But neither such energy conditions nor any other invariant properties of the spacetime play any role in \result.

\section{Past-eternal inflation}
\label{sec-pastei}

\subsection{Defining past-eternal inflation}

We are now almost ready to consider the implications of \result\ for past-eternal inflation.  But first, we must decide what past-eternal inflation actually {\em is}.  As a first step, let us distinguish between `past-eternal inflation' and `non-singular cosmology'.  The latter might be defined for present purposes as a model in which there exists at least one non-spacelike past-directed path with infinite proper time both to the past and to the future {\em at the classical level}, in standard GR. Such a path does {\em not} exist, for example, the the standard big-bang FLRW models if the strong energy condition applies.\footnote{This definition could be subdivided in various ways (note, for example, that this does not require that the universe be geodesically complete, or even complete to the past), but at the cost of great complication, as the study of cosmic censorship shows that careful and strict definitions can be maddeningly subtle.  Note also that this term is not intended to apply to models in which singularities are resolved or regulated away by quantum gravity or modifications to classical gravity.} Such non-singular models will be discussed in Sec.~\ref{sec-nonsing} so let us put them aside for now.

What, then, is past-eternal inflation?  First let us decide what it means for a point to be `inflating'.  I suggest that it entails two components: first, that there is `antigravity', i.e. the cosmic medium is dominated by a stress-energy component such that two test particles placed at rest with respect to each other will move away with time (essentially this requires violation of the strong energy condition).  Second, inflation must create \hotpocket s; that is, if one waits long enough (i.e. follows a wordline in the direction of increasing local time), the surrounding region will transition into one containing matter and radiation which then dilute with time in the coordinates in which they are locally homogeneous.  This definition of `inflation' would seem to cover all of the models I am aware of but not, for example, pure dS, nor any FLRW model with decelerated expansion.  Note that this definition is a relatively local one, and we can clearly have inflationary regions co-existing with non-inflating regions.

Finally, what is past-eternal inflation? This question turns out to be rather subtle. The basic idea should be that for any `time', the universe has inflation prior to that time. Denote, then, by `post-inflationary' a region that is not itself inflating, but for which every past-directed causal curve encounters an inflating region.  A spacetime might then be called past-eternally-inflating if at every time there are post-inflationary regions.  But what is a `time'?  Any spacelike surface can be a `time', at least locally. For `any time' to be well-defined, we need to foliate our full spacetime into spacelike surfaces; examples of such foliations of part or all of dS are given by the coordinate systems in Fig.~\ref{fig-dsreview}.  Not every spacetime admits a foliation, but ours does, or at least admits a set of foliations that cover the entire spacetime. Then, I shall call `perpetually inflating' a model in which  there exists a foliation of the full spacetime in which every equal-time surface contains post-inflationary regions.  (A more restrictive definition would hold that {\em every} foliation must entail this property.) In either sense, this would indicate that one cannot meaningfully point to a time before which where was no inflation.

What if a model does not admit a full foliation? We might then fall back on a weaker notion.  Suppose that there is a point $P$ in the spacetime such that for any proper time interval $\Delta \tau$, there exists a past-directed causal curve emanating from $P$ that encounters an inflating region a proper time $> \Delta \tau$ to the past. Roughly this means that there exists at least one infinitely old inflating worldline. I'll call this `ever-inflating', which also applies to any perpetually-inflating model.  A model that is perpetually-inflating (if it admits a foliation) or ever-inflating (if it does not) I will propose to call past-eternally inflating.

There are a few things to note about these definitions.  First, neither of them rules out singularities (or other geodesic incompleteness) to the past; we might wish to do I see no reason to tie this requirement into the definition of past-eternal when it is logically distinct. Second, both are more stringent definitions than that of `non-singular universe', because ever-inflating or perpetually-inflating implies non-singular, but the converse is not true (for example, Minkowski space is non-singular but nowhere inflating). Third, there is no obvious incompatibility between a \result\ and a model being ever-inflating, since as noted above, \result\ says nothing about past-directed geodesics that approach comoving ones.  Whether \result\ forbids perpetual inflation is less immediately clear, and is discussed next.

\subsection{Implications of the singularity theorems for past-eternal inflation}

Let us then finally turn to what singularity theorems mean for past-eternal (and particularly perpetual) inflation; their implications for non-singular cosmology will be discussed in the next section.

Drawing upon Sec.~\ref{sec-fvei} and Fig.~\ref{fig-dsreview}, we can start with a few candidate models of past-eternal inflation. In all three, the basic idea is to demand of the universe some type of time-translation symmetry. The first and oldest is the classic Steady-State cosmology.  In the model's original incarnation the universe was described by dS in the flat foliation.  Galaxies and smaller objects were distributed throughout this space, and a `C-field' was invoked to both supply negative pressure to drive the exponential expansion and also to generate new matter.  By construction, each constant-$t$ time-slice is (statistically) identical, so the model is fully time-translation-invariant. This was arguably past-eternal inflation, but not observationally viable.\footnote{An interesting wrinkle on this is a model in which the C-field is at very high-energy, but in certain regions becomes unstable so that the exponential expansion ceases and matter/radiation domination begins~\cite{1984JApA....5...67N}.  In this model the precise structure of the model was not spelled out, but it is conceptually almost identical to false-vacuum eternal inflation.}

Now consider false-vacuum eternal inflation as discussed in detail in Sec.~\ref{sec-fvei}, and take the inflating background metric to be, again, the flat foliation of dS.  Now imagine that at some time $t_0$, the universe is in pure false-vacuum.  After this, bubbles of true vacuum form, so that at each $t > t_0$ time-slice, there is a statistically homogeneous distribution of bubbles, with the largest being those formed near $t_0$.  As $t\rightarrow \infty$, the maximal bubble size diverges, and the bubble distribution approaches an asymptotic form.  Recall from the discussion of Sec.~\ref{sec-fvei} that although the inflating fraction approaches zero in this limit, that inflating regions {\em nonetheless exist}, forming a fractal of dimension less than three. Now, let us send $t_0 \rightarrow -\infty$; this just means that at any time $t$, the bubble distribution is exactly the one that was asymptotically approached as $t-t_0 \rightarrow \infty$ before.  Because in that case each time had both inflating and non-inflating regions, there is nothing inconsistent about assuming this state at each time $t$.  In this way, again, the universe becomes (statistically, on large-scales) time-translation invariant in the time $t$.  This model was described by Vilenkin in~\cite{Vilenkin:1992uf}.

A third model is the `cyclic/ekpyrotic universe' scenario~\cite{Steinhardt:2001vw,Steinhardt:2001st,Steinhardt:2002ih,Khoury:2003rt,Steinhardt:2004gk}, in which our 4D universe is a `brane' embedded within a higher-dimensional space, and periodically smashes into another such brane.  In each collision (which is singular at the level of 4D physics), energy is converted from the brane dynamics into radiation and matter, creating a\hotpocket\ that evolves as usual.  After a long time, the energy density becomes dominated by a tiny vacuum energy (presumably the one we observe now), and expands exponentially for a long time until the next brane-brane collision, which restarts the process.  This model is described in much more detail in another chapter of this volume.  Again by construction, each cycle is identical to the previous one and hence there need be no 'first' cycle.  Though its authors might take exception, this model is also eternal inflation by the definition I am using, since there is accelerated expansion followed by \hotpocket\ creation. (Another recent cyclic model, which relies on violations of the weak energy condition, is~\cite{Baum:2006nz}.)

\begin{figure}
\centering
\includegraphics[width=0.95\textwidth]{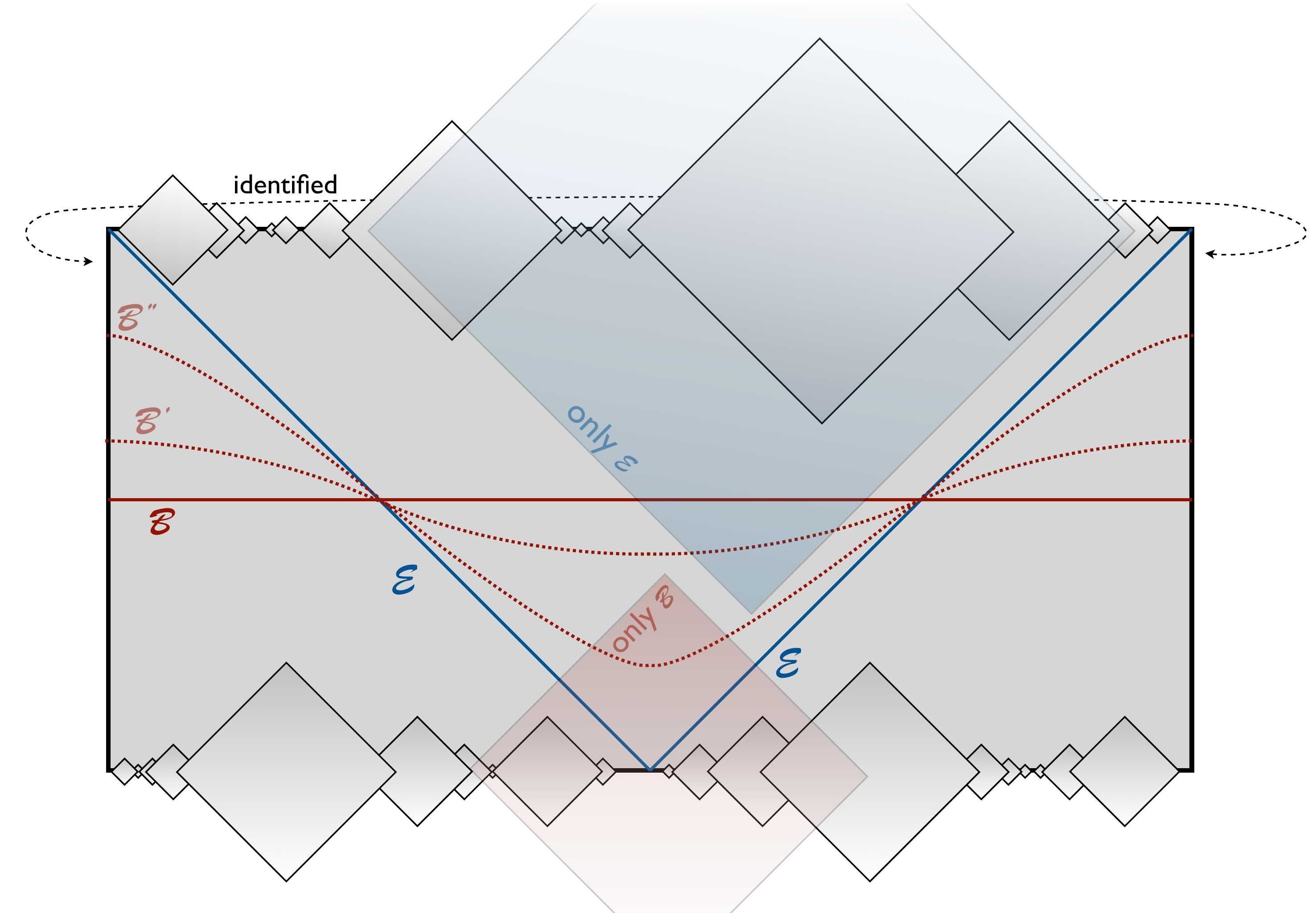}
\caption{A conformal `slice' of a picture of eternal inflation. The field is taken to be in the false vacuum on either ${\cal E}$ (for steady-state past-eternal inflation) or on ${\cal B}$ (for non-singular eternal inflation); the former is a certain limiting case of the latter.  Bubbles open away from these boundary-value surfaces, forming an infinite (and colliding) fractal on the past-and future- boundaries of the spacetime.
\label{fig-generalei}}
\end{figure}

Given three models, all of which are clearly `perpetually inflating', why is the viability of past-eternal inflation in question? In all three cases, the concern is the same, and is immediately apparent from examining Fig.~\ref{fig-dsreview} (left panel): just as when applying \result\ to dS, the issue is that the coordinates in which the spatial sections are flat, and in which the time-translation invariance holds, only cover half of dS.  That is, all three models are geodesically incomplete not due to any curvature singularity, but because their underlying spacetime is {\em extendible}, meaning that the manifold in question can be embedded as a submanifold in another manifold (in this case dS or something similar.)  This `edge', labeled \edge\ in Figs.~~\ref{fig-dsreview} and~\ref{fig-generalei} is exactly what \result\ points to when applied to these models. This raises two questions: first, are we bothered by this extendibility?  Is it a fatal flaw?  Second, if so, can we somehow fix it by performing the extension? 

Extendibility would normally be highly pathological, as one would naturally ask: `what would I see if I went to the edge?!'  If the spacetime is well-defined and finitely describable, it is much more natural to continue the manifold in some way, as is usually done upon discovering that an `end' of a spacetime is just a end of the coordinate patch (as for example in the static coordinatization of a Schwarzschild black hold outside of the horizon).  Yet in the cases in question, the edge is to the past of any observer, so we cannot imagine actually going there.  Moreover, if we imagine a material particle  `having come from there',  we come upon a new difficulty: as we go farther back in $t$, the particle's path was increasingly null with respect to comoving matter.  Thus a {\em physical} observer coming into the steady-state region would immediately encounter a blast of {\em infinitely} energetic particles (or bubble walls), violently distorting the spacetime and rendering the whole picture ill-defined.  That is, the steady-state description is simply {\em incompatible} with any material particles entering the steady-state region, and the formulation of the steady-state description is tantamount to disallowing any such particles (see~\cite{Aguirre:2001ks} for a more detailed version of this argument).  

For these reasons I think one can make a reasonable case that we could simply neglect the extendibility of the manifold.  But many will object to this, and I too personally find it disturbing.  Thus it is well-worth asking whether we could somehow modify the picture in a way that makes physical sense and also removes the distasteful extendibility of the region.

In some cases, at least, it appears that we can~\cite{Aguirre:2003ck}. Let us consider the above-described false-vacuum eternal inflation picture, under the assumption there are {\em no} transitions from the true vacuum to the false vacuum, and that bubble walls are perfectly null.  Then the nucleation point of any bubble must have no bubbles in its past lightcone.  Now we can ask: suppose we extend the manifold just a little bit, onto the null surface \edge.  What must the field be there?  Well, we know that the field arbitrarily close to \edge\  must be in the false vacuum (if not, we could just go to an earlier time closer to \edge, before the false-vacuum bubble in question formed).  Then if we simply demand continuity, then \edge\ must also be in the false vacuum.  

But this little extension buys a lot because \edge\ is actually `sort of' a Cauchy surface.  In particular, it is essentially a null initial value surface for the steady-state region. (The `essentially' is because the point $i^-$ labeled in Fig.~\ref{fig-dsreview} is not part of \edge; but as argued in~\cite{Aguirre:2003ck} this point has no effect on the fields in the steady-state region.)  But note that \edge\ is {\em also}, in exactly the same way, a boundary-condition surface for the region beyond \edge.  That is, if we extend the manifold, and demand continuity of the field onto \edge, then we also define what happens on the other side.  

So what happens?  Classically, nothing: the other side would be pure dS in the false-vacuum.  But that would violate quantum mechanics, since the false-vacuum is unstable to decay.  Yet if we allow a decay, we would apparently violate the boundary condition that \edge\ is in the false vacuum.  It seems we are at an impasse. However, there is a hidden assumption: that once nucleated, a bubble opens toward the top of the paper (or screen) you are looking at.  More precisely, the assumption is that bubbles open with a single, continuous time-orientation.  What if they don't?

In that case bubbles could also form past \edge, but open `downwards', away from \edge.  Then, as you can convince yourself with a bit of thought, the region is just like the steady-state region, just time-reversed.  Thus beyond \edge, there is simply {\em another} past- and future- eternally inflating universe.  Does this `reversal of the arrow of time' make sense?  I think it does: as argued at length in ~\cite{Aguirre:2001ks,Aguirre:2003ck}, and in the context of similar constructions in~\cite{Carroll:2004pn}, \edge\ would constitute a {\em minimum entropy} boundary condition.  Away from this surface, entropy must increase, and hence the arrow of time must point.  This idea will be further discussed below; for now the key point is that there is a well-defined model that is both past-eternally inflating by the definition I gave above, and also geodesically complete.  In fact it's hard for me to imagine a reasonable definition of `past eternal inflation' that would fail to apply to this model unless it is either specifically tailored to do so, or introduces some sort of `transcendent' basis for time-orientation beyond any standard notion of what physically determines the AOT.  Thus I think the widely-held inference that \result\ demonstrably disallows such models is simply false --  though whether any one such model is {\em true}, or free from any hidden fatal and irrepairable internal inconsistencies, is certainly an open question.

Is this the only answer?  Under the assumptions (two vacua, no upward transitions), the extension to \edge\ is fairly uniquely defined, so there is little wiggle room there.\footnote{As discussed at length in~\cite{Aguirre:2003ck}, there is a natural identification that physically equates the steady-state region with its copy, so that there is just one steady-state region, and geodesics reaching \edge\ simply emerge from another part of \edge.}  However, we might ask about more general potentials that allow upward-transitions, or support stochastic (or topological) eternal inflation; and we can ask about Cyclic models.

Let us first allow up-tunnelings from the true- to false-vacuum, as described in Sec.~\ref{sec-recycle} above.  Now the basic method employed above to fix the field value on \edge\ does not work, because we can{\em not} assume that the full past lightcone (up to \edge) of a bubble nucleation event is in the false-vacuum; a different approach to the problem is required. 

Consider $T=0$ in the closed foliation (see Fig.~\ref{fig-dsreview}), which is a Cauchy surface for dS. Let us place on this surface the boundary condition that the field is fixed in the false vacuum.\footnote{Quantum-mechanically, this can be done in the `Schrodinger' field-theory picture by defining wave functional $\Psi[\phi(\vec x)]$ to be nearly gaussian, and centered on the false vacuum at $t=0$ (see \cite{Rubakov:1999ir,Halliwell:1991ef}, so that the probability of any region having the field in the true vacuum at $t=0$ is very small.} The only bubble configuration consistent with this condition in one in which all bubbles open {\em away} from $T=0$, echoing the sort of time-symmetry of the steady-state model described above.  Now consider `boosting'\footnote{More precisely we can `boost' the surface by replacing it with the surface that $T=0$ maps to when the embedding Minkowski spacetime undergoes a boost} this $T=0$, $\phi=\phi_F$ surface to a new surface \boundary, as shown in see Fig.~\ref{fig-generalei}. Some points will remain on \boundary, but most $T=0$ points will now be either `before' or `after' it  (i.e. in a direction of decreasing or increasing $T$ from it), and thus may have bubbles running through them.  In fact, for greater and greater boosting, the volume available for nucleation between \boundary\ and any given point steadily increases -- but there is always a region very close to \boundary\ with no bubbles.\footnote{More specifically, the 3-volume of \boundary\ is independent of the boost,
so in the quantum sense it is equally easy to prevent a true-vacuum region from existing on any of them via the same gaussian wave functional centered on $\phi_F$.} We can then take the (infinite-boost) limit in which \boundary\ approaches \edge, and we have reproduced the steady-state model.  

What would happen if we put different boundary conditions on \boundary\ before we start boosting, such as seting $\phi=\phi_T$ on \boundary?  Despite some effort, I do not know, but I will speculate.  My intuition is that if we could clearly define the physical configuration corresponding to these (quantum) boundary conditions, then boost the surface on which we place them, we would find that the physical configuration is unchanged, i.e. the state would be dS-invariant.  Taking this to the infinite-boost limit, this would say that `steady-state recycling inflation' and `ground state' dS with a false vacuum are just the same thing: both involve a time-translation-invariant {\em and time-reversal invariant} mixture of true- and false-vacuum regions.  

To speculate further, I might suppose that this `taking the ground state' trick would work for any model (even those including topological or stochastic eternal inflation) in which the ground state is of positive vacuum energy, so that the other states may be reached from it.  But if the lowest vacuum is negative, it would fail; for example if the lowest vacuum were zero, we would probably\footnote{It seems to me that this depends on whether we allow acausal up-jumps; if we do, it seems that baby universes could be spawned even from bare Minkowski space.} just be defining stable Minkowski space.  In these cases we might still obtain eternal inflation, but only be defining boundary conditions on a null surface that are in some sense rather special, as in the steady-state model defined above.

What about `cyclic' models in which a relatively homogeneous universe undergoes cycles? First let us consider the `cyclic/ekpyrotic' model. This seems even harder to understand in terms of extending the spacetime.  The reason is that a worldline followed to the past toward the edge of the manifold passes through an infinite precession of cycles taking infinitesimal proper time.  Thus there is no well-defined sense of a smooth limit as the edge is approached (as in taking the limit as $t\rightarrow 0$ of $\sin (1/x)$.
Moreover there is an explicit breaking of dS-symmetry so we cannot use the trick of boosting \boundary\ and arguing that the result is independent of the boost.  Conceivably, we could imagine some state that `decays into' the oscillating state, then use the same method of boosting \boundary\ into \edge\ employed above.

As for the model of~\cite{Baum:2006nz} that employs violations of the weak energy condition to give bounces from contraction to expansion, I see several causes to question the viability of the model.
First, I am skeptical of their solution of the `entropy puzzle': their model is 4D and homogeneous, with a periodic scale factor $a$.  If I take a region of fixed comoving volume and follow it back in time (decreasing entropy in my comoving region) far enough it seems I must inevitably reach either zero or increasing entropy, both of which the authors reject.  Second, I am also extremely dubious of large-scale violations of the weak energy condition (as in this and in `big rip' cosmologies), because this condition is intimately tied to others at the foundations of physics, and significant violations wreak havoc on things most physicists hold very dear, such as the law of entropy increase.\footnote{For example, suppose it is possible to create a `lump' of material violating the weak energy condition.  It would have negative mass.  Now let me take some amount of matter, with entropy $S$, and form it into a black hole of mass $M$.  The thermodynamics of black holes is well-established, and holds that the black hole has an entropy $S_{BH}=4\pi GM^2/\hbar c$, and that $S_{BH} > S$ (the latter is known as the `generalized second law of thermodynamics'.  But now let me throw in my negative-mass lump.  Clearly I can decrease the black hole's mass, and hence entropy, until $S_{BH} < S$.  So I have violated the second law of thermodynamics.} Sir Arthur Eddington memorably wrote that  `...if your theory is found to be against the second law of thermodynamics I can give you no hope; there is nothing for it but to collapse in deepest humiliation'~\cite{eddington}.  This may be overly harsh, but I do think that theories flagrantly violating the weak energy condition bear a very heavy burden.

Leaving cyclic models and returning to inflation, we have seen that some cases of eternal inflation are `tractible' in that there is a well-defined dS `skeleton' that described the overall large-scale structure.  But in a very general potential that might drive false-vacuum topological {\em and} stochastic eternal inflation in a many-dimensional effective potential, one might despair of ever comprehending the cosmic large-scale structure.  The singularity theorems were designed to apply to even very general situations, but as we have seen, even in simple potentials with fairly well-understood physics they admit exceptions that allow steady-state eternal inflation.  What if we allow more general physics?  As noted above (and see e.g.,~\cite{Linde:2007fr}) even if {\em most} past-directed geodesics meet singularities, that may still square perfectly with past-eternal inflation, since most geodesics in future-eternal inflation do not keep inflating indefinitely either. This idea might underly a well-posed perpetually-inflating or ever-inflating (by the definitions I gave in Sec.~\ref{sec-pastei}) model, but nothing along these lines has been worked out.

On the other hand, the models described above avoid singularities by including a reversal of time's arrow along some worldlines.  The next section takes this idea -- that the singularity theorems are telling us that that many worldlines in inflating regions must have {\em either} classical singularities {\em or} time-reversal -- to its logical conclusion.

\section{Toward a nonsingular universe}
\label{sec-nonsing}

Might the singularity theorems point to either singularities {\em or} time-reversal along the past of any worldline?  A rough argument for this idea is fairly simple.  Consider the neighborhood (say a region of fixed comoving volume) of a wordline.  For an observer along that worldline to perceive an arrow of time (AOT) it must see local net entropy generation.  To the past, it seems entropy destruction.  In a finite neighborhood, there is finite entropy, so going far enough `to the past' the entropy must (a) asymptote to a constant, (b) start to increase again, or (c) become ill-defined as a singularity is encountered. 

It would be fun to try to make this argument precise, but in keeping with the spirit of this volume, what I will do instead is take an exploratory foray into the question of what happens if possibility (c) is eliminated, so that {\em there are no past singularities}. To do so I will propose a new sort of cosmic censorship.
	
\subsection{Consistent consmic censorship}

The cosmic censorship (CC) conjecture proposed by Penrose~\cite{Penrose:1969pc,Penrose:1979fz} has played a key role in the development of classical and quantum gravitational physics.  The literature contains many specific statements expressing two general forms of this conjecture. Roughly speaking, the {\em weak} form~\cite{Penrose:1969pc} asserts that singularities will not be visible to observers that can reach timelike or null infinity, but will instead be shielded from view by event horizons.  The {\em strong} form~\cite{Penrose:1979fz} asserts that {\em no} observers (including those inside horizons) will see naked singularities.

The initial motivation for CC was essentially pragmatic: if a naked singularity develops from some set of initial data, then the spacetime and its field content anywhere in the causal future of the singularity are unpredictable.  This assumption of classical predictability underlies many key theorems  in classical GR (see, e.g.,~\cite{1975lsss.book.....H,1984gere.book.....W}).  Moreover, it has the great virtue that some version of it may actually be true: although unproven in either basic form, it currently appears to admit no generic and generally agreed-upon counterexamples (see, e.g.,~\cite{Wald:1997wa,Brady:1998au} for reviews, but see also~\cite{Joshi:2002dt}). This, however, raises some very interesting questions, among which are: 
\begin{enumerate}

\item  If quantum gravity resolves classical singularities, enabling the prediction of the future of a naked singularity, why does nature work so very hard to prevent them from forming? That is, why is nature kind to classical relativists, but cruel to quantum relativists?

\item  If cosmic censorship is true in general, why would it admit one single, glaring exception filling the past lightcone of every observer, in the form of the `big-bang' cosmological singularity?

\end{enumerate}

These questions are rarely if ever addressed directly in the literature, but appear to be avoided through, in the first case, a tacit assumption that classical GR is self-consistent and well-posed and, in the second case, a assertion that the big-bang singularity is somehow qualitatively different from naked singularities that might develop from well-posed initial data.  I propose that these questions might be addressed by taking a more consistent view of cosmic censorship:

{\bf Consistent cosmic censorship (CCC):} {\em Without exception, no physical observer can physically observe a past singularity.}

This has two key points.  First, the discussion is of a {\em physical} observer making {\em physical} observations.  The reason for this is that such an observer experiences an AOT, inextricable from entropy increase, which defines what should be considered a past (rather than a future) singularity.  The hypothesis is, therefore that both (a) strong cosmic censorship holds (prohibiting timelike singularities), and also (b) entropy increases towards, rather than away from, spacelike or null singularities.  
In this view, singularities would represent a sort of gravitational equilibrium, which thermodynamically cannot be in the past of usual spacetime. (See~\cite{Banks:2006hy} for some arguments along these lines.)  Clearly, this is not a precise formation; it is meant as an idea, at the same level of precision as `physical observes will not see the entropy of a closed system decrease', which future work may or may not allow to be formulated more precisely.

The second key point is that the conjecture applies just as well to cosmology and rules out, in particular, the classic big-bang model.  What does that mean?  Of course, it does not mean that spacetimes violating CCC do not exist {\em mathematically}, any more than cosmic censorship means that solutions with naked singularities don't exist -- they of course do.  The idea is that nature refuses to make use of these solutions. I have a detailed set of careful and utterly convincing arguments in favor of CCC, but neither this article nor its footnotes can contain them.\footnote{I don't really have such arguments.  If I did, they might relate to the chain of reasoning that (a) assuming CC allows proof of the area theorems for black hole~\cite{1975lsss.book.....H} and (in some cases) cosmological~\cite{Davies:1987ti,Maeda:1997fh} horizon growth; (b) the area theorem and the identification of (one fourth) the horizon area with entropy allows a consistent generalized second law of thermodynamics; (c) the generalized second law underlies a consistent AOT for physical observers. Running this chain of reasoning backward, guaranteeing a consistent AOT for a physical observer requires CC in the sense that a candidate naked singularity potentially allows violations of the (generalized) second law, which in turn implies either an inconsistent AOT, or time `running backwards' so that the putative naked singularity is actually in the future rather than the past.}  So just as an interesting exercise, let us consider what it would mean for cosmology if we adopt this restriction.

\subsection{Cosmologies obeying the CCC}

Several CCC-consistent cosmologies have already been discussed in this paper, e.g. the steady-state eternal inflation model, and the `$\phi=\phi_F$ on \boundary' model as well as its $\phi=\phi_T$ variant.  In the first two, entropy effectively increases away from a low-entropy boundary condition placed on a spacelike or null surface; overall the universe obeys a sort of time-reversal symmetry.  In the last case, the boundary condition is {\em high} entropy and the universe can be considered as some sort of equilibrium, again including a sort of time-reversal symmetry (whether this falls afoul of the issues mentioned in Sec.~\ref{sec-recycle} is open for debate).  I find these models an amazingly neglected (until~\cite{Aguirre:2001ks,Aguirre:2003ck} and~\cite{Carroll:2004pn}) way to address the AOT: they allow an AOT without imposing time-asymmetry on the universe {\em or} forcing low-entropy boundary conditions both at the `beginning' and `end' of the universe (which seems to be almost universally the only sort of time-symmetric universe considered in the literature).

Another CCC-consistent example, without this symmetry, is the `emergent universe' of~\cite{Ellis:2002we}.  Here, the universe is a close Einstein static universe, in which the (repulsive) vacuum and (attractive) kinetic energies of a scalar field at $\phi \ll 0$ are perfectly balanced {\em forever} until the field `eventually' rolls into a true-vacuum well at (say) $\phi=0$.  (See Ch.~6 for more detail.) This is an interesting model but I am skeptical that it is self-consistent.  To balance for eternity, the kinetic and potential energies must {\em exactly} balance.  But considering this as a quantum system, this implies a $\delta-$function in (field) momentum space.  This, however, requires a completely flat wave functional in (field) position space, meaning that the potential `dip' at $\phi=0$ would always be probed by the wavefunctional, and moreover that one cannot say that the universe has any particular field value {\em or} is evolving. More philosophically, it is hard to imagine how nothing happens for an {\em infinite} time, then suddenly does! Nonetheless it is an interesting example that is consistent with both \result\ (because the integral of $H$ is finite), and with CCC.

What sort of CCC-consistent inflationary cosmology might we imagine in most generality?  Such a universe would have a mix of inflating and non-inflating regions, contain (only) spacelike singularities, and have regions with well-defined as well as ill-defined arrows-of-time.  Large areas of eternal inflation would spawn \hotpocket s, which might in turn eventually spawn new inflating regions.  Following any worldline back in (local) time from a \hotpocket, would eventually find itself in a region where the AOT is ill-defined, then `reverses' so that the worldline becomes future-directed; this (according to the CCC) would always occur before a singularity is reached.  Thermodynamically, the universe would, in both `time directions', be seeking to reach equilibrium in the form of big-crunch singularities or stable Minkowski space.  But the infinite degrees of freedom available to the Universe, as well as the structure of eternal inflation, would ensure that equilibrium is never globally achieved.  The universe, to the past and the future, but really always to the future, would be eternally interesting.

\section{Open issues}

I'll conclude this chapter with a list of some of the remaining questions in eternal inflation that I, at least, would like to have answers to.

\begin{itemize}

\item By what mechanism do `upward' transitions from low- to high- vacuum energy occur and how does this impact the picture of eternal inflation?

\item What do bubbles collisions look like and what does this mean for observations?

\item How exactly do we pose initial/boundary conditions for eternal inflation?  Are they somehow very special, or somehow equilibrium, or neither?

\item Is there a measure on eternally inflating spaces that is physically well-defined and well-motivated, paradox free, and does not lead to obviously incorrect predictions?

\item How generic is eternal inflation?  Are there reasonable models that reproduce the data and do not have eternal behavior?

\item Can the infinite volume of physical space in eternal inflation be squared with the `holographic complementarity' idea that dS has finitely many degrees of freedom?  In general what does holography say about eternal inflation?

\item What is the `deep reason' for cosmic censorship?  Does that deep reason also suggest Consistent Cosmic Censorship, or does it happily allow initial spacelike singularities like the classic big-bang?

\item How does the arrow of time arise in eternal inflation?  Does it ever reverse along a well-defined worldline? What would it mean for an arrow of time to be inconsistent for causally-connected observers? Are we missing something crucial in thinking of the universe as a closed system with some set of states and transitions between them?

\end{itemize}

\section{Appendix: energy conditions}

Singularity theorems (as for many general theorems in GR) in the literature often assume conditions obeyed by the energy-momentum tensor $T_{\alpha\beta}$.  Some common ones are defined here; see ~\cite{1975lsss.book.....H,1984gere.book.....W,2004sgig.book.....C}  for further discussion.
\begin{itemize}
\item {\em strong energy condition}: for all timelike vectors $t^\alpha$, $T_{\alpha\beta}t^\alpha t^\beta \ge {1\over 2}T^\mu_{\ \mu}t^\nu t_\nu$.  In terms of a perfect fluid of pressure $p$ and density $\rho$, this means $p+\rho \ge 0$ and $\rho+3p \ge 0$.

\item {\em weak energy condition}:  for all timelike vectors $t^\alpha$, $T_{\alpha\beta}t^\alpha t^\beta \ge 0$, or for a perfect fluid $\rho \ge 0$ and $\rho+p \ge 0$.

\item {\em null energy condition}: for all null vectors $n^\alpha$, $T_{\alpha\beta}n^\alpha n^\beta \ge 0$, or $\rho+p \ge 0$.

\item {\em dominant energy condition}: that the weak energy condition hold, plus $T_{\alpha\beta}T^\alpha_{\ \mu}t^\beta t^\mu \le 0$ (i.e. $T^{\alpha\beta}t_\alpha$ is non-spacelike), or that $\rho \ge |p|$.

\end{itemize}
%
% BibTeX users please use
\bibliographystyle{hunsrt}
\bibliography{../master_bibliography_physics}
%
% Non-BibTeX users please follow the syntax
% the syntax of "referenc.tex' for your own citations
%%%%%%%%%%%%%%%%%%%%%%%%%%%%%%%%%%%%%%%%%%%%%%%%%%%%%%%%%%%%%%%%%%%%%%  }

%%%%%%%%%%%%%%%%%%%%%%%%%%%%%%%%%%%%%%%%%%%%%%%%%%%%%%%%%%%%%%%%%%%%%%

\printindex
\end{document}